\documentclass{kais}
\usepackage{wrapfig,epsf}
\usepackage[agsmcite]{harvard}
\usepackage{hyperref}
\usepackage{graphicx}
\citationmode{abbr}
\renewcommand\harvardand{and}

\received{24 Feb 2016}
\revised{24 May 2018}
\accepted{31 Jul 2018}

\pubyear{2018}
\pagerange{\pageref{firstpage}--\pageref{lastpage}}
\volume{xxx}

\begin{document}
\label{firstpage}

\title{Compact and Efficient Representation of General Graph Databases\footnote{A preliminary partial version of this paper appeared in Proceedings of the Eighth Workshop on Mining and Learning with Graphs (MLG2010), pp. 18--25, 2010.}}

\author[S. {\'A}lvarez-Garc{\'i}a et al]{Sandra {\'A}lvarez-Garc{\'i}a$^1$, Borja Freire$^2$, Susana Ladra$^3$ and {\'O}scar Pedreira$^3$\\ $^1$Indra, A Coru{\~n}a, Spain; $^2$ Enxenio S.L., A Coru{\~n}a, Spain; $^3$ Universidade da Coru{\~n}a, \\CITIC, Database Laboratory, A Coru{\~n}a, Spain}
 
\maketitle

\begin{abstract}
In this paper, we propose a compact data structure to store labeled attributed graphs based on the $k^2$-tree, which is a very compact data structure designed to represent a simple directed graph. The idea we propose can be seen as an extension of the $k^2$-tree to support property graphs. In addition to the static approach, we also propose a dynamic version of the storage representation, which allows flexible schemas and insertion or deletion of data. We provide an implementation of a basic set of operations, which can be combined to form complex queries over these graphs with attributes. We evaluate the performance of our proposal with existing graph database systems and prove that our compact attributed graph representation obtains also competitive time results.
\end{abstract}

\begin{keywords}
Compression; Graph Databases; Property Graphs; Attributed Graphs; Compact Data Structures; Dynamic Graphs
\end{keywords}

\section{Introduction}

Graphs are a natural way for modeling data in such domains where the most relevant
information relies on the relationships between the entities. Some representative
examples are Web graphs \cite{raghavan03}, social networks \cite{dama10},
computational biology \cite{bohm00}, pattern recognition \cite{conte04},
chemical data analysis \cite{aggarwal10}, and geographic information systems, among others.
In the last years, many research
lines have emerged focusing on the analysis of data showing this graph nature.
Furthermore, in the Big Data Era, huge volumes of data are generated every day.
This information needs to be stored and processed efficiently in terms of space and
time. In this scenario, where graph mining involves complex analyses on
huge datasets, the design of new compact graph representations that can be
accessed efficiently has become an important research field.

In many cases, the graph models used to represent the relevant information for
a domain are simple directed graphs. Many compact and efficient proposals have
appeared to represent this kind of graph \cite{Jac89,BoVWFI,CKL+09,CN10,HN14,GLOUDS,Maneth2016}.
However, in many cases, this model is not
enough, because nodes and edges contain complex information that must be stored and accessed.
These domains where nodes and edges include a set of attributes (key/value) define a new model of
graphs, usually called {\em attributed graphs}
or {\em property graphs}. For this scenario, {\em graph database models} emerged to give a theoretical support to attributed graphs. These new models are characterized by representing the schema,
the data, the queries, and the results as a graph \cite{LMD14}. Built over those theoretical models,
many practical Graph Databases Engines have been proposed \cite{CAH12}.
DEX \cite{DEX,DEXRETRIEVAL} or Neo4j \cite{HHLD11} are two relevant examples.

Given the amount of information that those graph database engines have to
manage, it is important to focus on the design of efficient and compact structures
to represent attributed graphs, as it improves their management and querying.
In many domains, a plain representation of the graph may not fit in main memory,
and swapping can degrade the performance. Typical access patterns and navigation
over the graphs also make difficult an out-of-core processing of graph data.
Therefore, it will be important to compress and index the dataset, so it can be stored in main memory and
support efficiently the navigation operations, without the need of uncompressing
during the analysis of the dataset.
In this article, we propose a compact structure to store
attributed graphs, whose internal representation is based on the $k^2$-tree \cite{BLN14}, a static
structure designed to represent simple directed graphs (binary relations) in main
memory. Our goal is to study the possibility of extending this compact structure
to obtain a very succinct representation of attributed graphs that supports efficient graph operations and
access to the attributes of the nodes and edges of the graph. The $k^2$-tree has been
successfully extended in the past, for instance by proposing a dynamic variant that supported changes in the set of edges \cite{k2dyn}, or to support other types of data representation, such as temporal graphs \cite{Caro16,AGdBBNjda16}, RDF datasets \cite{AGdBBNjda16}, or raster data \cite{dBABNPspire13.3,k2raster}. For instance, RDF datasets can be considered ternary relations, thus, they can be decomposed into
a collection of binary relations, and represented with a variant of the $k^2$-tree that provides indexing capabilities in all the three dimensions \cite{AGdBBNjda16}. Raster data can be represented using also a collection of $k^2$-trees, one per each different value existing at the raster dataset \cite{dBABNPspire13.3}, or by generalizing the original method for integer values instead of bit values \cite{k2raster}. However, none of the previous works tried to extend the $k^2$-tree structure to represent general graphs, including labels and attributes.

The result of this work is the {\it Att$K^2$-tree}, a compact structure to store attributed
graphs based on the representation of binary relations in a very compact way using
the $k^2$-tree structure. In addition, we present a dynamic version, which allows changing both the schema and the data contained in the database. We denote it as  {\it dynAtt$K^2$-tree}.
We compare our proposals with other attributed graph representations in the
state of the art, obtaining the best space/time trade-off for basic query operations.

The paper is structured as follows. Section \ref{sec:stateGeneralGraphs} describes the most representative tools existing in the state of the art to manage attributed graphs. Section \ref{section:background} briefly reviews the $k^2$-tree, which is used in our proposal. In Section \ref{sec:GeneralProposal}, we present our compact data structure to store such graphs with attributes, that is, the Att$K^2$-tree. This section focuses on the physical storage. Section \ref{subsec:nav} presents the operations implemented in the Att$K^2$-tree.  Section \ref{sec:dynamic} presents the dynamic variant of our proposal, that is, dynAtt$K^2$-tree. Finally, Section \ref{sec:GeneralEvaluation} provides an experimental evaluation of the system using representative cases of study where we compare our proposal with other attributed graph representations in the state of the art.

\section{Systems for attributed graphs}\label{sec:stateGeneralGraphs}

Last years, graph database models have been proposed to represent attributed graphs.
These models specify the data, the queries, the results and, in many cases, even the schema of the graph as a graph \cite{GDMGU}. Many theoretical models and their corresponding query languages were proposed to represent and navigate graphs. Some examples are the \textit{Hypernode Model} \cite{HYPERNODE}, whose main feature is that nodes can be graphs by themselves, and GOOD (Graph Oriented Database Model)\cite{GOOD}, where data manipulation operations (insertions and deletions of nodes and edges, and clustering of nodes depending on some properties) are specified as graph transformations.

Built over those theoretical models, many practical {\em graph databases engines} have been proposed.  In this section, we describe some of the most relevant works in this area, including their internal data structures for
graph representation and processing. Some of the graph database systems reviewed in this section use storage data structures specifically designed for graphs, as it is the case of DEX, Neo4j, or HyperGraph, while others, such as SAP HANA Graph and SQLGraph, store graphs in relational database tables, or in a combination of relational databases with complementary stores.

\subsection{DEX}\label{sec:DEX}

DEX (now named as Sparksee\footnote{\url{http://sparsity-technologies.com/}}) \cite{DEX,DEXRETRIEVAL} is a graph database that efficiently stores and queries labeled and directed attributed multi-graphs. It keeps the graphs in secondary memory using different bitmaps. The graph model of DEX defines labeled nodes and directed edges where extra information is associated to each node and edge, represented as a list of attributes. Therefore a graph in DEX is defined as $G=(V,E,L,T,H,\{A_1\dots A_n\})$ where:

\begin{itemize}
 \item $V$ is the set of node keys.
 \item $E$ is the set of edge keys.
 \item $L$ is a key-value list that includes each key node (or key edge) and its label.
 \item $T$ and $H$ are key-value lists that associate each edge key to the keys of its corresponding source (tail, $T$) and target (head, $H$) nodes respectively.
 \item Each $A_i$ in $\{A_1\dots A_n\}$ represents a different attribute. Nodes and edges of the graph can take values for some of these attributes.
\end{itemize}

\begin{figure}
  \centering
      \includegraphics[width=0.6\textwidth]{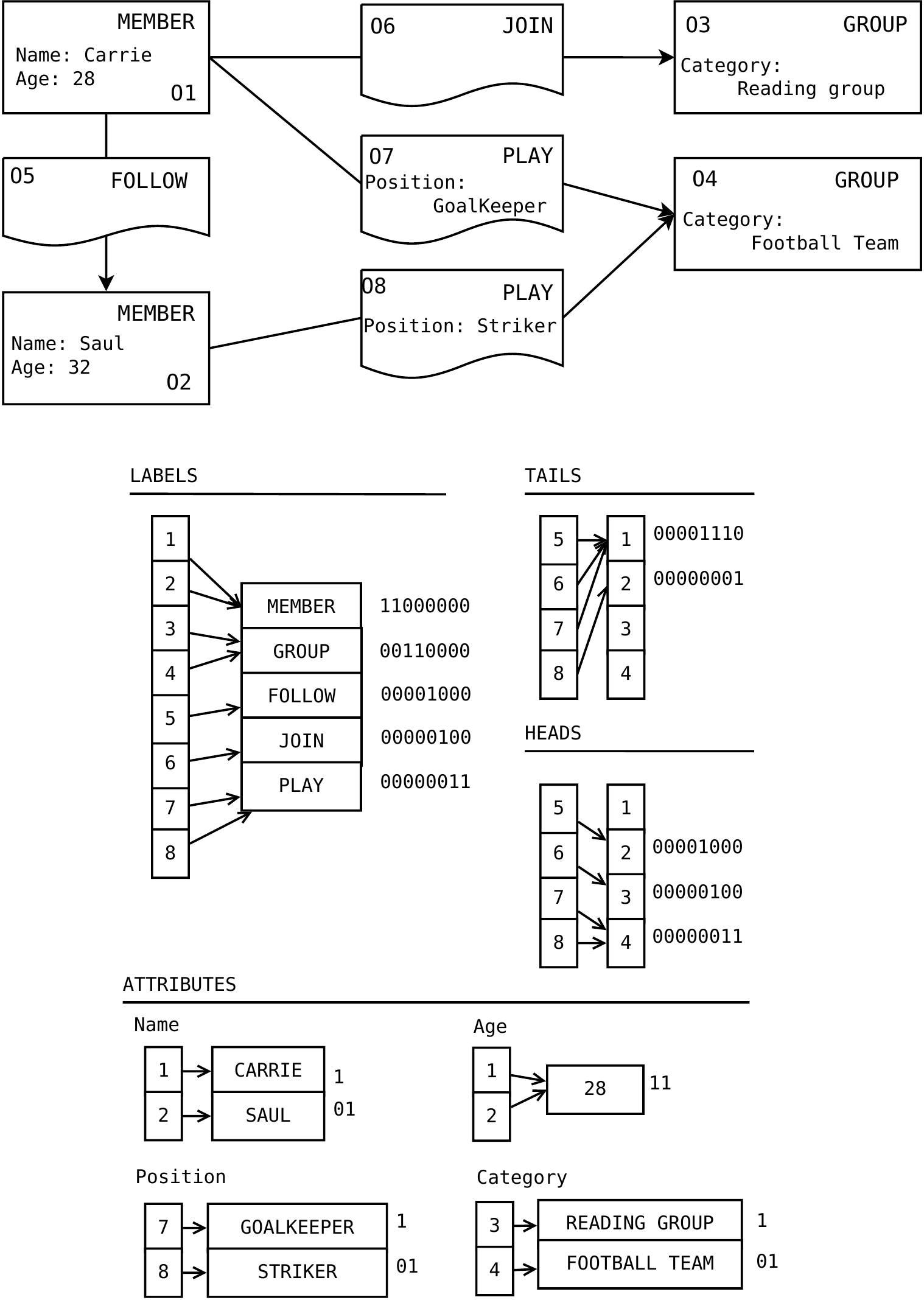}
  \caption{Example of labeled attributed graph (top) and its DEX internal representation (bottom).}
  \label{fig:dex}
\end{figure}

DEX represents this graph model through a set of bitmaps. Figure \ref{fig:dex} shows an example of a graph internal representation in DEX. The top of the figure shows a graph for a social network where two members (\textit{Carrie} and \textit{Saul}) have joined different groups (a \textit{reading group} and a \textit{football team}). Users can be related to each other through a \textit{follow} relationship, and users can participate in the groups through two different relationships (\textit{join} for a \textit{reading group} and \textit{play} for a \textit{football team}) described by different attributes (for example, the {\em position} attribute in the case of the {\em play} relationships). All the elements of the graph (nodes and edges) have an associated object identifier. For instance, the object identifier of the member \textit{Carrie} is \textit{O1}.

The bottom part of the figure shows the bitmaps used in DEX to represent this graph. Labels storage is shown on the left. A map (implemented using a $B+$-tree) associates each object identifier (node or edge) to its corresponding label value. For instance, the object identifier \textit{O3} is related to the label \textit{Group} (since it represents the \textit{Reading group}). Each label has an associated compressed bitmap that contains as many \textit{bits} as objects in the graph. For a given label, a \textit{one} in position $i$ means that the object $O_i$ is labeled with that label. However, in practice, bitmaps are only stored until the position of the last \textit{one}. The $B+$-tree and these compressed bitmaps compose a double mapping. The purpose of storing this double mapping is to support bidirectional navigation. The first map is used to obtain the label of each object identifier. The bitmaps of each label answer the opposite query: given a label (for instance, \textit{member}), obtaining the objects with that label is performed by retrieving the \textit{ones} in the corresponding bitmap. The bitmap for \textit{member} is $[11000000]$, meaning that \textit{O1} and \textit{O2} are objects with the \textit{member} label. The remainder information for the graph is managed in a similar way: a double map is used to represent the tails of the edges, another one represents the heads, and finally one map per attribute is stored.

The main purpose of this internal structure is to provide bidirectional access. That is, the data for a given node can be recovered using the maps indexed by its identifier. On the other hand, given a label or an attribute value, finding the nodes or edges with this label or value is performed by checking its corresponding bitmap and recovering the positions with a value \textit{one}. Direct and reverse neighbor nodes are recovered by using the bitmaps \textit{head} and \textit{tail}.

DEX query engine is built over this internal representation. Its core implements a small set of primitives, and more complex queries are built on top of them.

\subsection{Neo4j}\label{sec:neo4j}

Neo4j\footnote{\url{http://neo4j.org/}} is an open-source graph database that supports the storage and query of labeled directed attributed graphs. A graph in Neo4j can be defined as $G=(N,E)$, where $N$ and $E$ are the sets of nodes and edges respectively. Each node is a pair $n_i=(L_i,P_i)$, where $L_i$ is the set of labels and $P_i$ is the set of properties (or attributes) of the node. Labels of a node can be seen as tags. They can be used to define constraints over a group of nodes, to represent temporary states of nodes or, in general, to define a target group of nodes over which an operation will be performed. A property $p_j \in P_i$ is a key-value pair, where the value can be a primitive (the typical primitive types of any programming language like Boolean, Integer and String are supported) or a list of elements from one of these primitive types. An edge of a graph in Neo4j is defined as $e_i=(n_j,n_k,t_i,P_i)$, where $n_j$ and $n_k$ are the nodes linked through this edge, $t_i$ is the label of the edge and $P_i$ is the set of properties the edge contains (equivalent to the properties of the nodes). Neo4j defines its own query and update language, Cypher, which is a declarative language, where data is obtained by pattern matching.

Neo4j uses native structures for storing graph data, that is, it does not rely on traditional relational storage structures. As explained in \cite{RWE13}, the different parts of the graph, nodes, edges, and properties are stored in different store files. For example, nodes are stored in fixed-record files (with a length of 9 bytes), and the position of each node in the file is given by the node identifier. Similarly, edges are stored in fixed-record files (33 bytes per record, in this case). This file stores for each edge the identifiers of the source and target nodes, a pointer to the edge type, and pointers to the next and previous edges of the source and target nodes, which allow for a faster query processing. The properties of both nodes and edges are stored as key-value pairs in a property store.

\subsection{HyperGraphDB}
\begin{figure}
  \centering
      \includegraphics[width=0.7\textwidth]{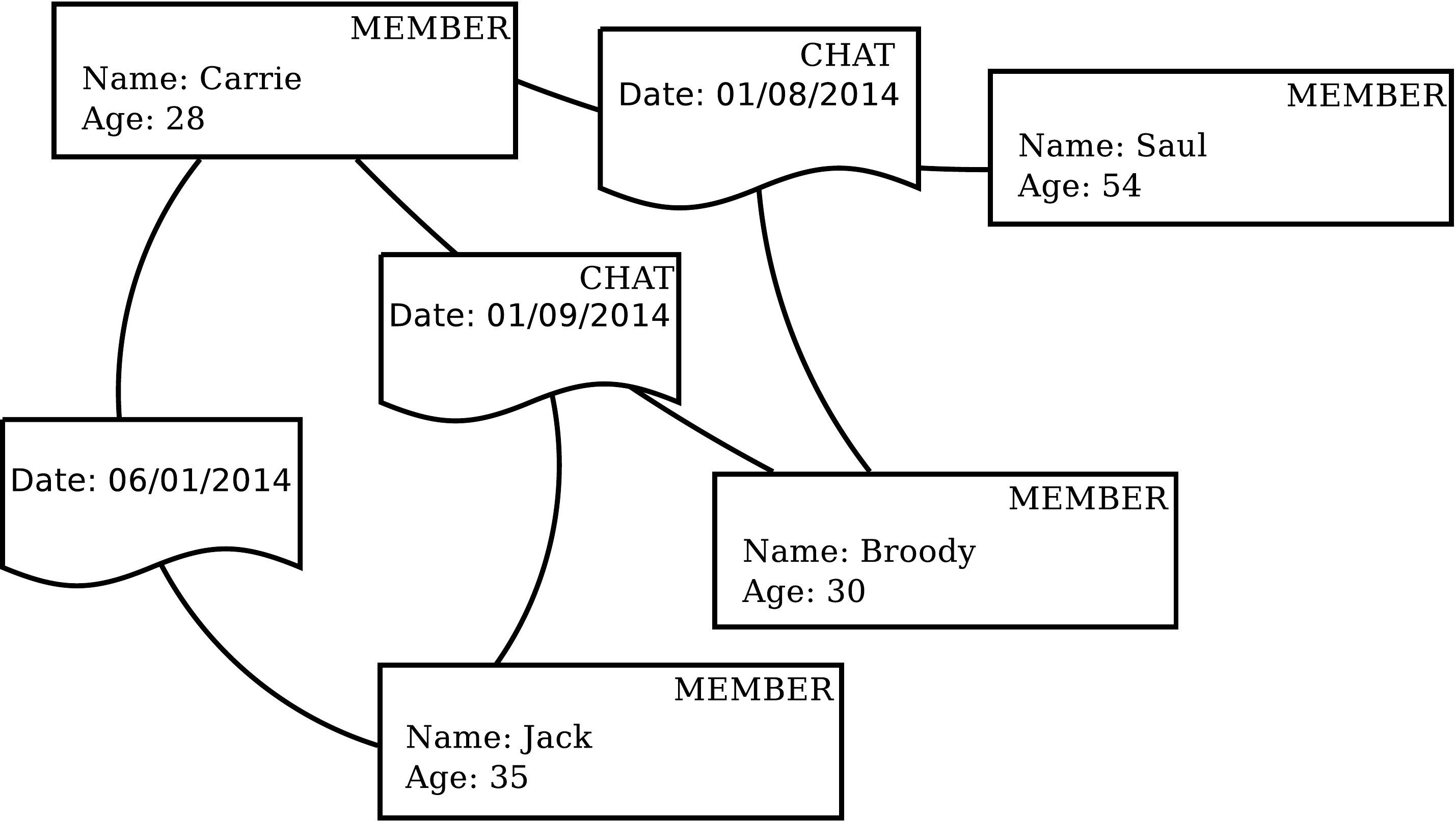}
  \caption{An example of hypergraph model including $n$-ary edges}
  \label{fig:hypergraph}
\end{figure}

HyperGraphDB\footnote{\url{http://hypergraphdb.org/}} \cite{HYPERGRAPH} is a graph database based on the Hypergraph model designed mostly for knowledge management, AI and semantic web projects. It supports a hypergraph $HG=(N,E)$, where $N$ is the set of nodes and $E$ is the set of edges, also known as links. The definition of links here is different from regular graphs, as links point to an arbitrary number of elements instead of just two, and links can be pointed to by other links as well. 
The hypergraph defines a more expressive structure that can be useful to model domains where more than two entities are usually related. For instance, each conversation of people in an online chat program could be modeled using this hypergraph model as a link that relates the participants in the conversation. Figure \ref{fig:hypergraph} shows an example of conversations along the time. For instance, the chat conversation in $01/08/2014$ is represented as a link that involves 3 members.

HyperGraphDB uses an {\em atom} as the basic unit of representation. It contains a \textit{typed value} and a \textit{target set} composed of a set of atoms. Atoms can be nodes or links. A node is just an object value that does not point to anything else. Thus, the number of atoms in the target set of a node is $0$. On the other hand, a link has at least one atom associated with it.

The storage of HyperGraphDB is basically distributed in two layers. The primitive storage layer includes
the information of the links (for each edge identifier the set of related identifiers is stored) and the data (the RAW value corresponding to each identifier). These data are stored in two associative arrays, implemented as key-value stores in BerkeleyDB. In addition to these two associative arrays, the system creates indexes on the data and allows the users to create additional indexes to speed up particular queries.

Over this primitive layer, the model layer manages the type system, the querying engine and some optimization facilities like caching and indexing. This layer manages the storage of the atoms, with their type, value and target set.

The HyperGraphDB query engine provides two different ways of specifying the query: an API to define standard graph traversals and a SQL-style language where a set of constraints over the required atoms are set.

\subsection{SAP HANA Graph}\label{sec:hana}
SAP HANA\footnote{\url{https://www.sap.com/products/hana.html}} is an in-memory database management system developed by SAP. An important feature of HANA is that it includes a module for general graph database management, called HANA Graph \cite{HANAGraph}. It allows managing graph databases in which edges are directed, two given nodes can be connected by different edges, and both nodes and edges can have attributes consisting of an attribute name, a data type, and a value. 

As described by \citeasnoun{HANAGraph}, both nodes and edges must have an identifying attribute, called vertex key and edge key respectively, and edges must have two additional attributes, the source and target nodes they connect. Graphs in HANA are stored in two relations or views, one storing the nodes of the graph, and another one storing the edges. In addition to the identifying keys and the source and target attributes, any other attributes of nodes or edges will be stored as columns in the corresponding table. HANA Graph provides a SQL-based language called GraphScript, a procedural domain-specific language \cite{Graphscript} that allows the users to easily manage nodes and edges, and to implement typical graph processing algorithms.

\subsection{SQLGraph}
SQLGraph \cite{SQLGraph} is a proposal for property graph storage that relies on existing database management systems rather than on specific data structures for storing and processing the graph data. SQLGraph follows a hybrid approach that combines relational databases in combination with JSON stores.
The adjacency information of the graph is stored in relations for the outgoing and incoming edges of the graph. However, this data is not represented plainly in the relations. Instead, hashing techniques presented in \citeasnoun{Bornea2013} are applied in order to improve query performance. The attributes of both nodes and edges are stored in additional JSON stores. In addition, these JSON stores keep a copy of the adjacency information of each edge, since this can improve the performance of certaing graph queries. The authors show in \citeasnoun{SQLGraph} how queries expressed using Gremlin \cite{Gremlin}, a procedural graph traversal language, can be translated into SQL queries on the relational database that stores the graph.

\subsection{Plain relational representation of graphs}
In addition to the systems we have reviewed in this section, attributed graph databases can also be created and managed in any relational database management system. A possible approach for this is analogous to the one used by HANA Graph, that is, storing the nodes and edges of the graph in two relations. The nodes relation would have a primary key and one column for each possible attribute present in any node. The edges table would have a primary key, mandatory source and target columns, and a column for each possible attribute present in an edge. Creating indexes on the primary keys of both relations, and on the source and target columns of the edges relation would improve the performance for basic graph operations, like obtaining the neighbors of a given node. Additional indexes created on certain attributes could also improve performance on queries involving those attributes.
Another possibility for storing and processing graph databases in relational databases would be creating different relations for the different node or edge types. That is, if our graph contains three types of nodes, where all the nodes of the same type share the same attributes, we could create three relations, one for each node type. The same design would apply for edges. The advantage of this approach is that we would not have so many null values in the tables, although the query performance could be worse, depending on the structure of the graph and the specific queries of interest.
In any case, although this is an option for representing attributed graph databases, the performance, both in terms of space and query times, would depend on the specific relational structure used to represent the graph.

\subsection{Other systems}

In the past years, many other graph database systems have emerged. They are focused on managing large amounts of data in a very efficient way. OrientDB\footnote{http://www.orientechnologies.com/orientdb/} is a good example, which is document and graph oriented, implemented in Java and uses SQL as query language. In addition, many proposals were designed to work in distributed environments. Titan\footnote{http://thinkaurelius.github.io/titan/}, Giraph\footnote{http://giraph.apache.org/}, or Pregel \cite{PREGEL} are just some examples.

\section{Background: the $k^2$-tree}
\label{section:background}

Before presenting our proposal in detail, in this section, we will briefly explain the data structure it is based on, that is,
the $k^2$-tree \cite{BLN14}. The $k^2$-tree was originally proposed for compressing Web graphs,
but can be used to represent any simple directed graph (that is, without attributes, labels, nor multiple edges linking two given nodes), and more generally,
to represent any binary relation. The $k^2$-tree is a compact tree structure created from the adjacency matrix of the graph by taking advantage of large empty areas in the matrix (that is, large areas in which there are no $1$s). It achieves a very compact
space representation of the graph, and supports efficient navigation, both forwards and backwards,
without the need of decompressing. In addition, this representation supports some navigation possibilities not supported
by other graph compression techniques, such as range queries over the adjacency matrix (that is, obtaining any submatrix), which are necessary in our proposal.

\subsection{Data structure and construction}

\begin{figure}
\centering
\begin{center}
\begin{tabular}{c}
\includegraphics[width=0.35\textwidth]{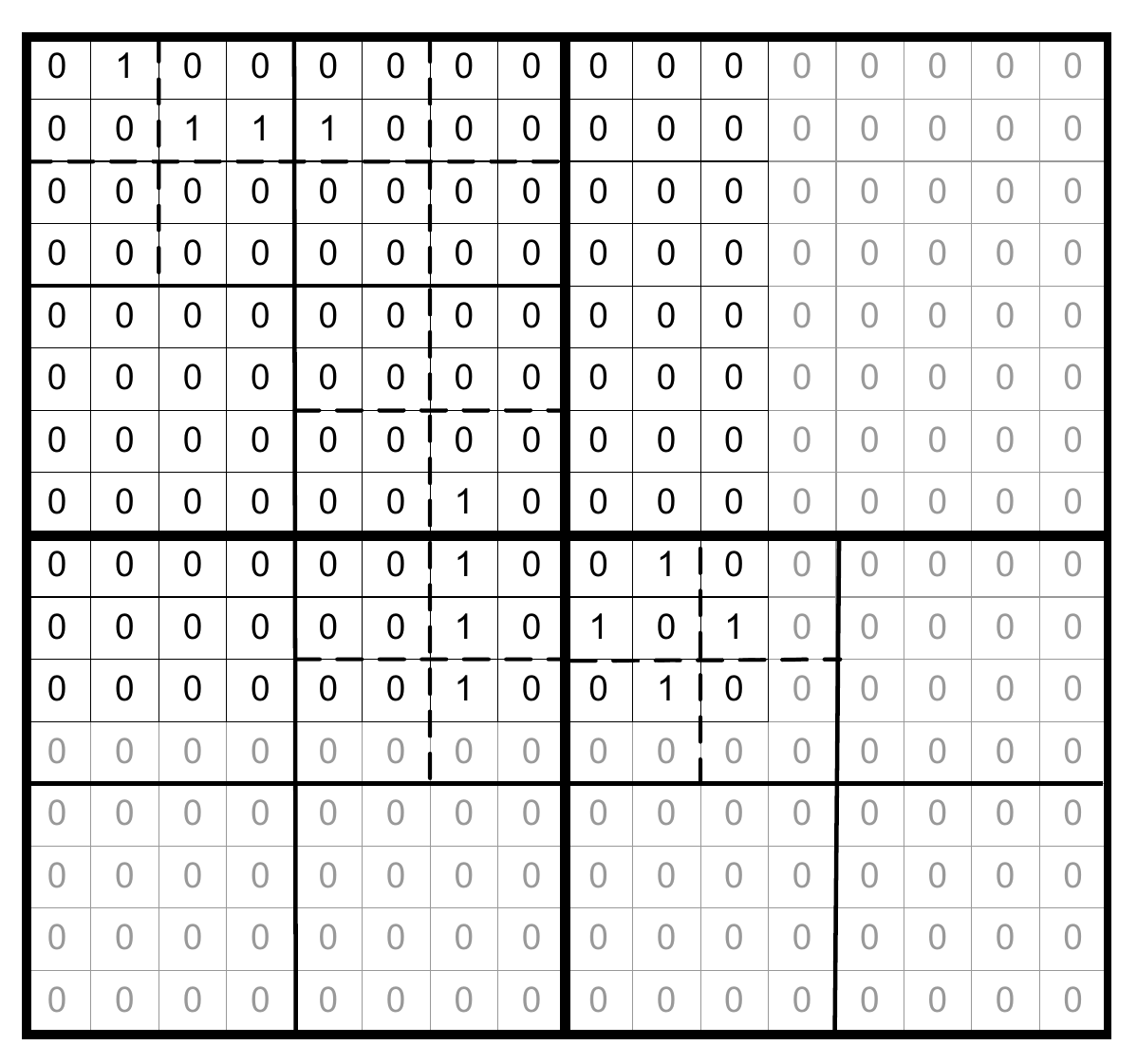}
\\
\includegraphics[width=0.45\textwidth]{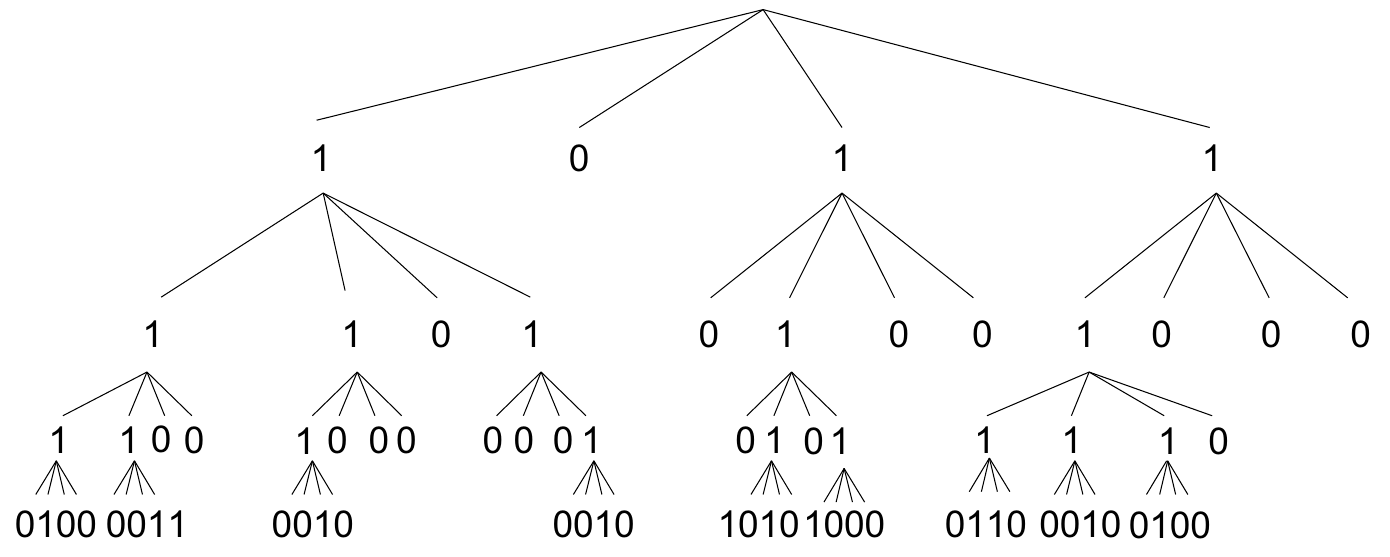}
\end{tabular}
\end{center}
\caption{Example of subdivision of an adjacency matrix (top)
and resulting $k^2$-tree (bottom), for $k=2$.}
\label{fig:options}
%and $k=4$ (right).}
\end{figure}

The $k^2$-tree is a tree-shaped representation of the adjacency matrix of the graph, built by recursively partitioning the adjacency matrix. The adjacency matrix of a graph with $n$ nodes is a square matrix $\{A\}$ of size $n \times n$, where row $i$ and column $i$ correspond to the $i^{th}$ node of the graph. The cell $A_{ij}$ is $1$ if there is a direct edge from node $i$ to
node $j$, and $0$ otherwise. The $k^2$-tree obtains its best performance when there are large matrix areas containing only $0$s, since these areas are represented with just one bit in the tree.

The construction of the $k^2$-tree consists in a recursive partition of the adjacency matrix
following a MX-Quadtree strategy \cite{Sam05}. This partition is conceptually represented using a non-balanced $k^2$-ary tree, in which each node contains one bit, and $k^2$ children.
Without loosing generality, let suppose that $n$ is a power of $k$. In a first level, $A$ is partitioned into $k^2$ submatrices of the same size. The root of the tree contains one child for each submatrix. The bit corresponding to each submatrix is $1$ if the submatrix contains some $1$, or $0$ otherwise (that is, if the submatrix is an ``empty'' area of the adjacency matrix). Then, the partitioning procedure is recursively applied to each non-empty submatrix until we reach empty submatrices, or the cells of $A$ containing a $1$, which are represented in the leaves of the tree. If $n$ is not a power of $k$, the adjacency matrix can be artificially expanded so its size is the next power of $k$, filling the new cells with $0s$. Since the $k^2$-tree represents empty areas of the matrix with one bit, this expansion of the adjacency matrix will not affect the final result significantly.

Figure \ref{fig:options} shows an example of how the $k^2$-tree is built, with $k=2$. In this example, the size of $A$ is $n=11$, but $A$ has been expanded by adding $5$ rows and columns, so the size of the new matrix is $16$. In a first partition of $A$, only the up-right submatrix is empty and, therefore, represented with a $0$. The rest of the submatrices are partitioned following the same schema, until we reach the leaves. The resulting tree is an alternative, more compact, representation of the adjacency matrix that allows us to obtain the value of a cell, row, column, or range of the adjacency matrix.

In order to avoid the use of pointers, the $k^2$-tree is represented in a very compact way using just two bit arrays: $T$ (tree) and $L$ (leaves).
$T$ stores all the bits of the $k^2$-tree except those in the last level. The bits are placed following a
levelwise traversal: first the $k^2$ bits of the children of the root node, then the bits of
the second level, and so on. $L$ stores the last level of the tree.
The $k^2$-tree uses an auxiliary structure that supports $rank$\footnote{$rank(B,i)$ computes the number of ones
that are set up in bitmap B until position $i$.} operations over $T$ in constant time,
which will be required to navigate the compact representation of the tree, that is, to traverse down from
the position of a node in $T$ to the start position of its children.

\subsection{Navigation}

To find the direct (reverse) neighbors of a node in the graph, the $k^2$-tree needs to locate
which cells in the row (column) of the adjacency matrix corresponding to that node have a
$1$. If we want to search for direct (reverse)
neighbors in a $k^2$-tree, we go down
through $k$ children forming a row (column) inside the matrix,
those submatrices that overlap with the row (column)
of the node of the query. This top-down traverse can be performed efficiently
over the compact representation of the tree, that is, the concatenation of
the two bit arrays and the auxiliary structure. Detailed algorithms for these operations, as well as for range queries, are thoroughly described by  \citeasnoun{BLN14}.

While alternative compressed graph representations
are limited to retrieving the direct, and sometimes the reverse, neighbors of
a given node, the $k^2$-tree representation allows for more sophisticated forms of
retrieval. 
First, in order to determine whether a given node $u$ has a direct edge to
a given node $v$, most compressed (and even some classical) graph representations have no
choice but to extract all the neighbors of $u$ (or a significant part
of them) and see if $v$ is in the set. The $k^2$-tree technique can answer such query in
efficient time, by descending to exactly one child at each level of the
tree.
A second interesting operation is to find the direct neighbors of node $u$
that are within a {\em range} of nodes $[v_1,v_2]$ (similarly, the reverse
neighbors of $v$ that are within a range $[u_1,u_2]$).
Yet a third operation of interest is to find all the edges from a range of
nodes $[u_1,u_2]$ to another $[v_1,v_2]$. 

\section{Our proposal: Att$K^2$-tree}\label{sec:GeneralProposal}

In this section, we propose our system, called {\em Attributed $k^2$-tree (Att$K^2$-tree)}, to efficiently store and process attributed graphs. The representation of the graphs is based on the $k^2$-tree, which, as seen, is a static data structure designed to work in main memory. Therefore, we propose an in-main-memory compact attributed graph representation designed to be used in contexts where big amounts of static data need to be intensively queried.

\subsection{Graph model supported by Att$K^2$-tree}

\begin{figure}
  \centering
      \includegraphics[width=0.68\textwidth]{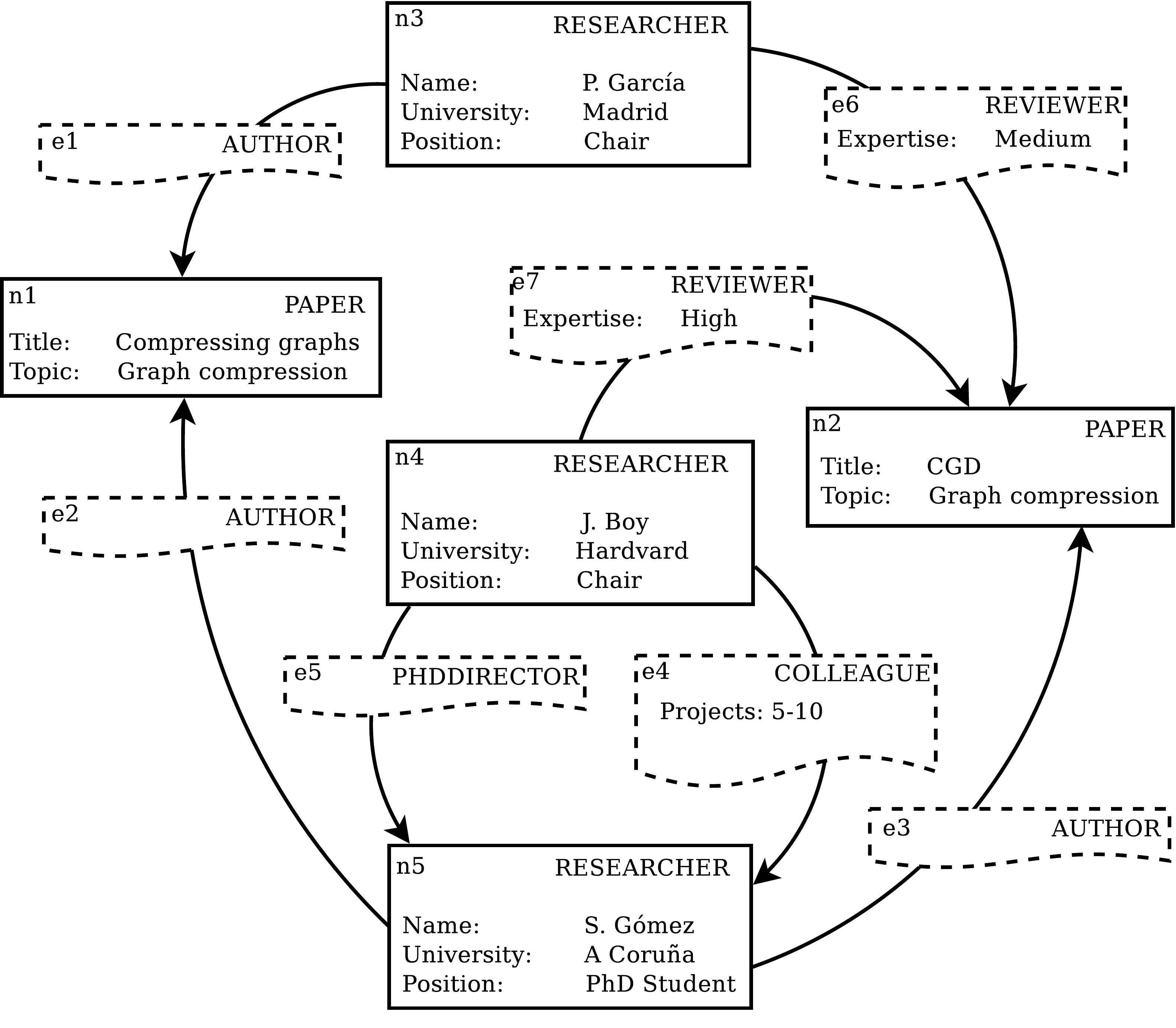}
  \caption{Example of a labeled, directed, attributed multigraph.}
  \label{fig:attGraph}
\end{figure}

We described several attributed graph systems in Section \ref{sec:stateGeneralGraphs}. All of them are based on specific attributed graph models, presenting only small differences between them. Therefore, before describing the internal representation of our structure, we first consider the features of the attributed graph model the Att$K^2$-tree supports. Figure \ref{fig:attGraph} shows an example of the attributed graphs supported by the Att$K^2$-tree. This graph represents a research network, including information about publications, authors, authorship, reviews, collaborations, and thesis supervision. Researchers and papers are modeled as nodes in the graph, and the different collaborations between them are reflected as edges (like thesis guidance and supervision, or collaboration in a research project). Researchers are related to the papers they {\em authored} through edges. Researchers can also be related with a paper through a \textit{review} relation. This graph is an example of the data that could provide support to an application that detects conflicts of interest in order to assign reviewers to papers.

The graph model of the Att$K^2$-tree presents the following properties:

\begin{itemize}
 \item{Directed graph}: the edges of the graph will be directed, meaning that they distinguish between origin and target node. Figure \ref{fig:attGraph} shows how edges from the graph explicitly identify their origin and target nodes. For instance, edge $e_1$ represents the authorship of a paper, having researcher $n_3$ as origin and paper $n_1$ as target of the edge. As usual, an edge of an undirected graph could be also represented in this model by using two directed edges (in opposed directions) between the two nodes it relates.

 \item{Attributed graph}: the attributes or properties are the most meaningful  characteristic of general graphs. Many approaches can be followed to define the attributes of an element, including complex data types, range domains, and another constraints over each attribute. However, we define a more simplistic conception of attributes. Each node and each edge of the graph is described through a set of attribute-value pairs. Values are not restricted to a domain or a data type. They can take any value which will be managed as plain text. In Figure \ref{fig:attGraph}, we can observe that edge $e_6$ takes value \textit{Medium} for attribute \textit{Expertise}.

 \item{Labeled graph}: several definitions can be considered for a labeled graph. As it was described in Section \ref{sec:stateGeneralGraphs}, DEX considers that each component of the graph (nodes and edges) contains a unique label (or main value) that identifies the kind of element it belongs to. On the other hand, labels in Neo4j are considered as tags, supporting the definition of multiple tags for each element. We consider, in line with DEX, that each element of the graph (node or edge) has just one label, which we call \textit{type}. This type determines the attributes that an element of that type can contain. In that sense, the label and the list of valid attributes for each label compose a \textit{schema} which can be very helpful to work with domains with structured data. Figure \ref{fig:attGraph} shows the label of each node and edge. In this example, two different labels are contemplated for the nodes. That is, that graph has only two node types: \textit{Researcher} and \textit{Paper}. The label determines the attributes that describe each node. Researchers are described through the attributes \textit{Name}, \textit{University} (where they work) and \textit{Position} in that university. On the other hand, papers are described through the {\em Title} and the main \textit{Topic} of the paper. Edges of the graph can be labeled with \textit{Author}, \textit{PhDDirector}, \textit{Reviewer} and \textit{Colleague}. \textit{Colleague} relates two researchers that have collaborated in some research project. To summarize, labels are used to identify the type of a node or edge.

 \item{Multigraph}: Att$K^2$-tree does not constrain the number of edges connecting two different nodes, thus, many edges with the same origin and the same target can be defined. This characteristic is useful to represent in a natural way contexts where several kinds of relationships can be established between two nodes. Figure \ref{fig:attGraph} shows an example of multigraph, where nodes $n_4$ and $n_5$ are related through two edges ($e_4$ and $e_5$) representing relationships with different nature (\textit{PhDDirector} and \textit{Colleague}) between those nodes. This multi-edge nature, combined with the labeled and attributed properties, makes this model very expressive. Therefore, Att$K^2$-tree can fit with the structural properties of many real graphs.

\end{itemize}

\subsubsection{Formal definition}\label{sec:attgmodel}

A formal definition of a labeled, directed, and attributed, multi graph \textit{(LDAM)} is represented as a 10-tuple $G=(L_N,L_E,N,E,R,L_A,S_N,S_E,A_N,A_E)$, where:

\begin{itemize}
	\item \textbf{$L_N$} is the set of possible labels that the nodes of the graph can take. For the graph in Figure \ref{fig:attGraph}, $L_N=\{\textit{Paper},\textit{Researcher}\}$, so there are two node types in this example.
	\item \textbf{$L_E$} is the set of possible labels that the edges of the graph can take. Regarding to the same example, $L_E=\{\textit{Author},\textit{Colleague},\textit{PhDDirector},\textit{Reviewer}\}$, so there are four types of edge.
	\item \textbf{$N=\{(n_i,l_j)\}$} is the set of nodes, being $n_i \in 1 \ldots |N|$ a numeric identifier of the node and $l_j \in L_N$ the label of the node. In the example, the set of nodes is composed of
	$$N=\{(1,\textit{Paper}), (2,\textit{Paper}),(3,\textit{Researcher}), (4,\textit{Researcher}), (5,\textit{Researcher})\}.$$
	\item \textbf{$E=\{(e_i,l_j)\}$} is the set of edges, where $e_i \in 1 \ldots |E|$ is the identifier of the edge and $l_j \in L_E$ is its label. The set of edges of the graph in Figure \ref{fig:attGraph} is 
	$$E=\{(1,\textit{Author}), (2,\textit{Author}), (3,\textit{Author}),(4,\textit{Colleague}),$$ $$(5,\textit{PhDDirector}),(6,\textit{Reviewer}), (7,\textit{Reviewer})\}.$$
	\item \textbf{$R$} contains the relations between the nodes, that is, the origin and target nodes of the edges of the graph. Each element of $R$ is a triple  $(e_i,o_i,t_i)$, where $e_i$ is the edge identifier, $o_i$ is the identifier of the origin node of the edge, and $t_i$ is the identifier of the target node of the edge. It is easy to note that, by definition, $|R|=|E|$. In the example:
	$$R=\{(1,3,1),(2,5,1),(3,5,2),(4,4,5),(5,4,5),(6,3,2),(7,4,2)\}.$$
	
	\item \textbf{$L_A=\{a_i\}$} is the set of the different attribute labels of the graph. In other words, $L_A$ is the union of all different attributes that describe the nodes and the edges of the graph. In the example,
	$$L_A=\{\textit{Title},\textit{Topic},\textit{Name},\textit{University},\textit{Position},\textit{Projects},\textit{Expertise}\}.$$
	
	\item \textbf{$S_N=\{sn_i\}$} is the set of schemas for the types of the nodes. Each element of $S_N$ defines the set of attributes that can be used for a given node type. Each element $sn_i$ of the schema is represented as a pair $(l_i,\{a_j\})$ that associates a set of attributes $\{a_j\} \subseteq L_A$ to a given node type $l_i \in L_N$, where the node label $l_i \in L_N$  has associated a set of attributes $a_j \in L_A$ defined by that node type. Note that an attribute is not exclusive of a node type. In other words, several node types can be described through the same attribute. Table \ref{table:example-schema} shows the schema of the nodes for the graph in Figure \ref{fig:attGraph}.
\begin{table}
  \caption{Nodes schema.}
\begin{center}	
\begin{tabular}{|l||l|}
  \hline
  Label ($l_i$) & Attributes($\{a_j\}$) \\
  \hline
  \hline
  \textit{Paper} & \textit{\{Title,Topic\}} \\
  \textit{Researcher} & \textit{\{Name,University,Position\}} \\
  \hline
\end{tabular}		
\end{center}
  \label{table:example-schema}
\end{table}
	
	\item \textbf{$S_E=\{se_i\}$} is the set of schemas for the types of the edges, in a completely analogous way to $S_N$. Each element of $S_E$ defines a valid schema for an edge type. Each $se_i=(l_i,\{a_j\})$ is a pair where $l_i \in L_E$ is the corresponding edge label and $\{a_j\}$ is the set of valid attributes for each edge type. Table \ref{example-schema2} shows the schema of the nodes for the graph in Figure \ref{fig:attGraph}.
\begin{table}
  \caption{Edges schema.}
\begin{center}	
\begin{tabular}{|l||l|}
  \hline
  Label ($l_i$) & Attributes ($\{a_j\}$) \\
  \hline
  \hline
  \textit{Author}      & \textit{\{ \}} \\
  \textit{Colleague}   & \textit{\{Projects\}} \\
  \textit{PhDDirector} & \textit{\{ \}} \\
  \textit{Reviewer}    & \textit{\{Expertise\}} \\
  \hline
\end{tabular}
\end{center}	
\label{example-schema2}
\end{table}	

	\item \textbf{$N_A=\{ (n_i,a_j,v_k) \}$} defines the properties of the nodes. It is a set of triples, where each triple defines the value $v_k$ that the node  $n_i \in 1 \ldots |N|$ takes for the attribute $a_j \in L_A$. Note that a triple $(n_i,a_j,v_k)$ is valid in a data source if $ \exists \ l_m | (n_i,l_m) \in N \wedge (l_m,\{\ldots a_j \ldots\})  \in S_N$. That is, a node can only take a value for an attribute included in its schema, given by its node type. For instance, the set of triples describing the properties of node $n_3$ in Figure \ref{fig:attGraph} are shown in Table \ref{table:example-attributes}.

\begin{table}
    \caption{Attributes for node $n_3$.}
\begin{center}	
\begin{tabular}{|l||l||l|}
  \hline
  Node Identifier ($n_i$) & Attribute ($a_j$) & Value ($v_k$) \\
  \hline
  \hline
  3 & \textit{Name} & \textit{P. Garc{\'i}a} \\
  3 & \textit{University} & \textit{Madrid} \\
  3 & \textit{Position} & \textit{Lecturer} \\
  \hline
\end{tabular}	
\end{center}
\label{table:example-attributes}
\end{table}

	\item \textbf{$E_A=\{ (e_i,a_j,v_k) \}$} describes the properties of the edges (analogously to $N_A$). As an example, the triple describing edge $e_6$ is provided in Table \ref{example-schema2}.

\begin{table}
  \caption{Attributes for edge $e_6$.}
\begin{center}	
\begin{tabular}{|l||l||l|}
  \hline
  Edge Identifier ($e_i$) & Attribute ($a_j$) & Value ($v_k$) \\
  \hline
  \hline
  6 & \textit{Expertise} & \textit{Medium} \\
  \hline
\end{tabular}	
\end{center}
\label{table:example-attributes2}
\end{table}	

\end{itemize}

Next, we detail the internal representation of Att$K^2$-tree designed to support the graph model presented in this section.

\subsection{Data structure}\label{sec:attdatastr}

The Att$K^2$-tree stores a directed, attributed, and labeled multi-graph by using binary relations represented with $k^2$-tree structures. It is a compressed solution composed of a set of $k^2$-trees and some additional data structures. The Att$K^2$-tree represents the graph with three components: the schema of the data, the data included in the nodes and the edges and, finally, the relations between the elements of the graph topology. Next, we present the three components of the Att$K^2$-tree.

\subsubsection{Schema}

The schema of the graph comprises the set of valid node labels (types) and edge labels (types), and the valid attributes for each of them, that is, the attributes that can be used for each node or edge type. The schema component works as an index for the other components of the Att$K^2$-tree. The elements of the graph model $L_N,L_E,N,E,S_N,S_E$ are stored in this schema layer. Figure \ref{fig:CGDSchem} on the left shows the schema storage for the graph of the example. It is composed of:

\begin{figure}
  \centering
      \includegraphics[width=0.68\textwidth]{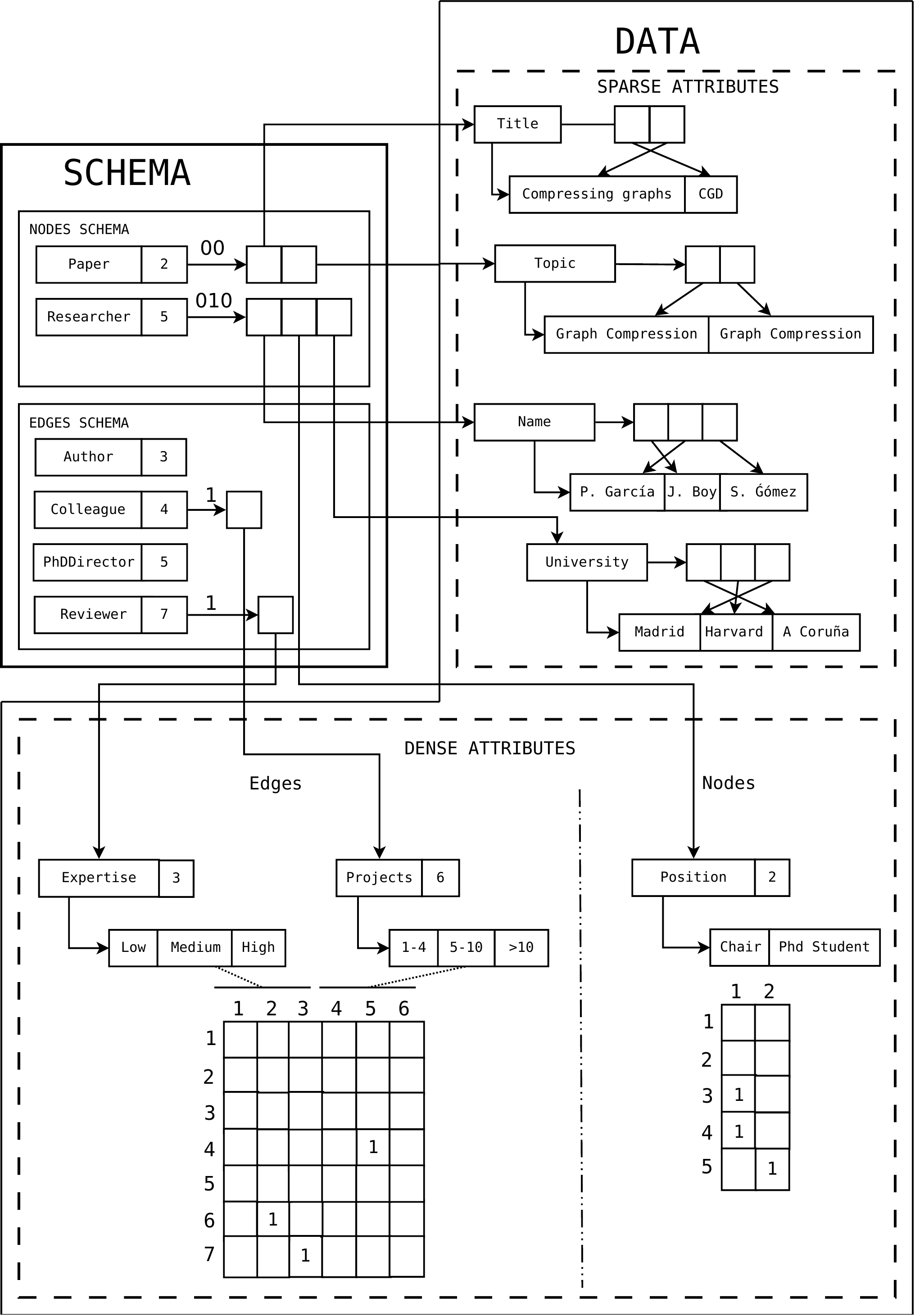}
  \caption{Internal representation of Schema and Data subsystems in Att$K^2$-tree.}
  \label{fig:CGDSchem}
\end{figure}

\begin{itemize}

\item \textbf{Nodes schema}: the nodes schema keeps the valid node labels, and the label each node of the graph has. To keep this part of the graph compact, node labels are sorted lexicographically, and node identifiers are assigned sequentially to the nodes of each label. That is, the first $m_1$ nodes with the first type in the schema will have identifiers from $1$ to $m_1$. The $m_2$ nodes from the second node type will have a range of identifiers from $m_1+1$ to $m_1+m_2$ and so on. Each entry of the nodes schema will store the highest node identifier with this label. Figure \ref{fig:CGDSchem} shows the nodes schema for the graph at Figure \ref{fig:attGraph}. It has two entries: \textit{Paper} (having 2 as the highest identifier) and  \textit{Researcher} (with limit $5$). That means that the nodes with identifiers in the range from $1$ to $2$ are papers, while the nodes with identifiers from $3$ to $5$ are researchers. Each label also points to its valid attributes in the data subsystem, and includes a bit array indicating whether each of its attribute is dense or sparse (further explained in the next subsection).

\item \textbf{Edges schema}: a table storing the edges schema is implemented in the same way as the nodes schema. Therefore, the edge identifiers will also be ordered by type. In that way, given an edge identifier, its corresponding type can be computed by performing a binary search over the entries in the schema. For instance, to recover the type of edge $6$ in Figure \ref{fig:CGDSchem}, a binary search over the upper limits of each edge label is performed, until reaching the range 5--7 that includes it, concluding that the node type of edge $6$ is \textit{Reviewer}.
\end{itemize}

The schema layer is the starting point of the internal representation of the graph in the Att$K^2$-tree, providing indexed access to the other two layers. It is used to retrieve the ranges of identifiers for a label, and to recover the label for a given identifier. It also stores references to the valid properties for a given label.

\subsubsection{Data}

Part of the data component of the graph contains the attribute values for the nodes and edges of the graph. It stores the values that each element of the graph (node or edge) takes for each valid attribute according to its type and the schema of the graph. Each different attribute can be represented in two different ways depending on the frequency distribution of its values. One type corresponds to the \textit{dense attributes}, where many nodes or edges of the graph share the same value for that attribute. In opposition to the dense attributes, in \textit{sparse attributes} nodes or edges usually take different values for that attribute. Titles, URLs, or identifiers are common examples of sparse attributes whereas age or nationality are examples of dense attributes. These two types of attributes will have a different internal representation in the Att$K^2$-tree:

\begin{itemize}

\item \textbf{Sparse attributes}: attributes for which the graph elements (nodes and edges) usually take different values will be stored as a list indexed by element identifier. This list is double-indexed: in addition to the implicit index by element identifier, there is an additional index to  maintain the entries in lexicographical ordering. This additional index is used to recover the elements taking a specific value by a binary search. Figure \ref{fig:CGDSchem} (top-right) shows four sparse attributes: \textit{Title, Topic, Name, University}. For instance, \textit{Name} is a valid attribute for the node type \textit{Researcher}. The values in the list are sorted by node identifier. The first element of this list \textit{(P. Garc{\'i}a)} is the value that the first researcher ($n_3$) takes for attribute \textit{Name}. Given a node $(n_i,l_j)$, its value for a sparse attribute will be in the position $i-\textit{limit}+1$, where \textit{limit} represents the lowest node identifier of the type $l_j$. The additional index provides support to perform a binary search over the attribute values. We can see in the example that the first element of this additional index in the attribute \textit{Name} points to \textit{J. Boy}, the first element of the list in a lexicographical order.

\item \textbf{Dense attributes}: all dense attributes of the graph are stored in two $k^2$-trees: a $k^2$-tree for the dense attributes of the nodes, and another $k^2$-tree for the dense attributes of the edges. The $k^2$-tree for the dense edge attributes is built as follows (the $k^2$-tree for the dense node attributes is built in the same way): each dense attribute $A_i$ can be seen as a binary relationship between the $|E|$ edges and the set of different values that those edges take for that attribute. These relationships can be represented in consecutive columns of the adjacency matrix. Rows of the adjacency matrix represent the edges of the graph, ordered by their identifiers. Columns will represent the possible different values of each attribute. Each group of consecutive columns represents the different values for an attribute. A $1$ in a cell $(i,j)$ of this adjacency matrix means that the edge with identifier $i$ takes the value $j$ for the attribute located in this range of columns. This adjacency matrix is represented using a $k^2$-tree. An additional structure stores, for each attribute, the block of columns that correspond to this attribute, storing again the upper limit for each attribute, and the specific values that represent each column.

Figure \ref{fig:CGDSchem} (bottom-right) shows the representation of the dense attributes. The adjacency matrix for the nodes is on the right, where the $5$ rows represent the $5$ nodes of the graph. The adjacency matrix on the left contains a row for each one of the $7$ edges. On the top of this adjacency matrix, the meaning of each column is specified by several lists. The attribute \textit{Expertise} contains three possible values \textit{(Low, Medium, High)}. This attribute includes the index $3$, which indicates that its three columns end at column 3 of the global adjacency matrix. Then, the cell $(6,2)$, which contains a $1$, means that the edge $e_6$ takes the value \textit{Medium} for attribute \textit{Expertise}. On the other hand, attribute \textit{Projects}, which is specified in the Schema as a valid attribute for the edge type \textit{Colleague}, contains three possible values which end at column 6 of the global adjacency matrix. In that way, the $1$ in cell $(4,5)$ means that $e_4$ takes value \textit{5--10} for attribute \textit{Projects}. Note that in some regions of this matrix, due to the schema constraints, no ones can appear. For instance, the matrix between the rows 1--3 and the columns 1--3 is empty because label \textit{Author} does not have attribute \textit{Expertise}, according to the schema.

\end{itemize}
Note that for some attributes, the choice of representing them as a sparse or dense attribute could be not obvious. A possible criteria could be based on the number of different values regarding to the number of elements taking a value for that attribute. 

To indicate if an attribute is dense or sparse, Att$K^2$-tree stores in its schema a bit array for each node/edge type, containing as many bits as valid attributes it has. A 0 value indicates that the attribute is sparse, and a 1 value indicates that the attribute is dense. For instance, the bit array corresponding to node label \textit{Researcher} is $010$, which indicates that the first attribute (\textit{Name}) is sparse, the second attribute (\textit{Projects}) is dense, and the third attribute ({\em University}) is sparse.

\subsubsection{Relations}

The third component of the Att$K^2$-tree stores the \textit{Relations}, that is, the different edges that connect the nodes of the graph. We store these relations with a $k^2$-tree, which needs to be extended to store the edge identifiers corresponding to each connection.

\begin{figure}
  \centering
      \includegraphics[width=0.7\textwidth]{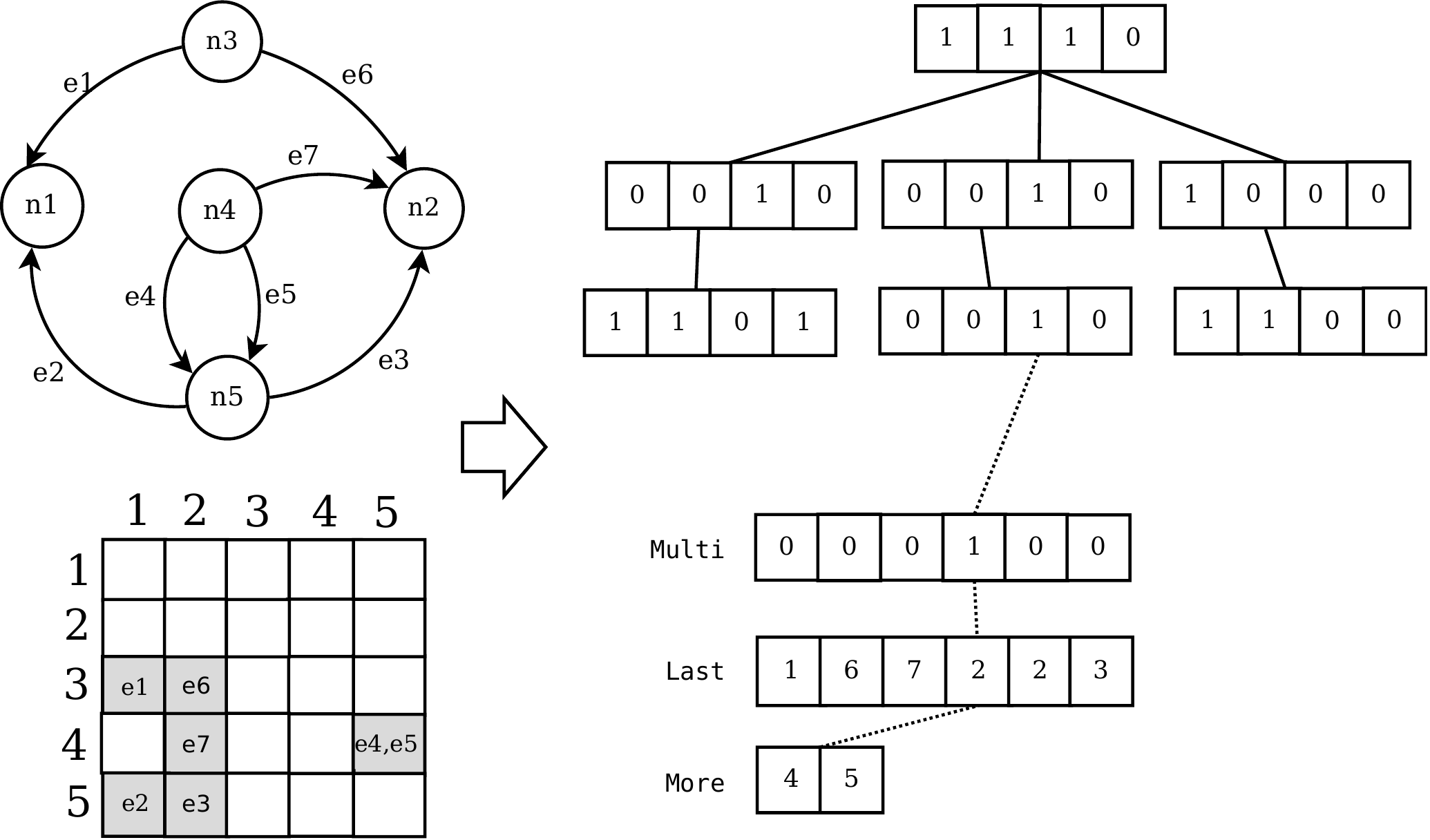}
  \caption{Relations component in the Att$K^2$-tree.}
  \label{fig:CGDRelations}
\end{figure}

The $k^2$-tree represents, in a very compact way, simple graphs that can be represented by an adjacency matrix. A \textit{one} in cell $(i,j)$ shows the existence of an edge from the node $i$ pointing to the node $j$. However, additional information is needed to store the relations in Att$K^2$-tree. First of all, each $one$ of the matrix has to be related to its edge identifier, which is used as pointer to the data layer (for instance, to recover the attributes of that connection). On the other hand, the Att$K^2$-tree supports multi-graphs. This means that more than one edge can relate a pair of nodes. So, several edges can be represented in the same cell of the adjacency matrix. Figure \ref{fig:CGDRelations} shows the relationships of the example and the corresponding adjacency matrix containing those edge identifiers. For instance, cell $(4,5)$ contains two edge identifiers because two different edges connect $n_4$ and $n_5$ in the original graph $(e_4, e_5)$.

The relationships in Att$K^2$-tree are stored with the original $k^2$-tree and some additional data structures to represent multi-edges and to trace their edge identifiers. Figure \ref{fig:CGDRelations} shows the structure we call multi-edge $k^2$-tree, which is composed of the following elements:

\begin{itemize}
\item \textbf{$k^2$-tree}: a $k^2$-tree is built to represent a binary relation among nodes in which two nodes are related if at least one edge connects them in the original graph. 
    It is a standard $k^2$-tree except for the fact that in this case, the bitmap of the last level also needs an additional structure to perform rank operations. By allowing rank operations, it is possible to compute the relative position of a bit with value $1$ in the last level of the tree, so it can be used as an index to the \textit{Multi} bitmap (which will be explained in the next paragraph). For instance, the last level of the tree in  Figure \ref{fig:CGDRelations} contains 6 \textit{ones}. If we perform a rank operation over the bitmap until the $7$-th position, we have that this is the fourth leaf of the $k^2$-tree, and we can check the fourth value of \textit{Multi} bitmap to check whether this leaf represents one or several edges connecting the same pair of nodes.

\item \textbf{Multi}: is a bitmap that stores, for each \textit{one} element of the leaf level, whether it is a multiple edge or it is representing only one edge. Therefore, $\textit{Multi}[i]$ will have value \textit{one} if the $i$-th one of the $k^2$-tree is clustering multiple edges. In the example, only the fourth position of the bitmap \textit{Multi} contains a \textit{one} value (clustering edges $e_4$ and $e_5$). This information is used to read the next array (\textit{Last}).

\item \textbf{Last}: this array stores the last edge identifier of each \textit{one} of the $k^2$-tree. For the $i$-th one of the leaf level, if it is a single edge (that is, if \textit{Multi}$[i]=0$) then $\textit{Last}[i]$ contains the identifier of that unique edge. Otherwise, when the $i$-th one is a multiple edge, $Last[i]$ represents the position in \textit{More} array, where the last edge is located. In the example of the Figure \ref{fig:CGDRelations}, {\em Multi}$[1]=0 \wedge \textit{Last}[1]=1$, so the last (and unique) edge (corresponding to cell $(3,1)$) has identifier $1$. On the other hand, $\textit{Multi}[4]=1 \wedge \textit{Last}[4]=2$ so the last edge of cell $(4,5)$ will be in $\textit{More}[1]$. It is possible to obtain the position of the first edge in array \textit{More}, as it will be the following to the last edge of the previous multi-edge. Thus, we first need to compute where the previous multiedge is in bitmap {\em Multi}, which can be denoted $p$, and then access to $\textit{Last}[p+1]$. Position $p$ is computed using rank and select operations\footnote{$select(B,i)$ obtains the position in $B$ of the $i$-th 1.} over {\em Multi}, more concretely, $p=select(rank(\textit{Multi},i)-1)$. Thus, in case that $\textit{Multi}[i]=1$, edge identifiers are located in $\textit{More}[b], \dots,More[e]$, with $b=Last[select(rank(Multi,i)-1)+1]$ and $e=Last[i]$. 

\item \textbf{More}: This array contains the identifiers of the multi-edges and it is indexed by the \textit{Last} and \textit{Multi} arrays. Figure \ref{fig:CGDRelations} shows the two identifiers for the only multi-edge of the example: $e_4$ and $e_5$, corresponding with cell $(4,5)$. 

\end{itemize}

The three layers Schema, Data and Relations compose the internal representation of Att$K^2$-tree, used to store  directed, attributed, labeled multi-graphs. These structures, based on the usage of $k^2$-trees, were designed to provide a compressed representation of attributed graphs, which could be accessed by basic queries. The next section presents the navigation over the internal representation of Att$K^2$-tree.

\section{Navigation and operations}\label{subsec:nav}

In this section we present the query API of Att$K^2$-tree composed by a set of basic operations over attributed graphs. This API aims to provide a basis for the construction of more complex queries. Our API contains 12 operations, which can be classified according to the layer of Att$K^2$-tree they imply.

\subsection{Operations over the schema}

Some of the basic operations in the Att$K^2$-tree work with the labels (types) of the graph.

\begin{itemize}

\item \textbf{Retrieval of labels}. The operation $\textit{Get}\{\textit{Node}|\textit{Edge}\}\textit{Types}$ returns the different labels of the nodes or the edges of the graph. According to the internal representation of the Att$K^2$-tree, it is trivially implemented by recovering all entries of the Nodes Schema (or the Edges Schema). In the  graph of the example, \textit{GetNodeTypes} returns the labels \textit{Paper} and \textit{Researcher}. On the other hand, \textit{GetEdgeTypes} returns the labels \textit{Author, Reviewer, PhDDirector, Colleague}.

\item \textbf{Filter by type}. $\textit{Scan}\{\textit{Nodes}|\textit{Edges}\}(\textit{Type})$ returns the nodes or the edges of a given type. Taking into account that the identifiers were allocated according to the type of the elements, this operation becomes quite straightforward. For instance, the operation $\textit{ScanNodes}(``\textit{Researcher}")$ is implemented by performing a binary search over the labels in the Nodes Schema, showed in Figure \ref{fig:CGDSchem}. When the entry $2$ is retrieved, the upper limit of the range of identifiers with label \textit{Researcher} is obtained, with value 5. The lower limit of the range is retrieved from the previous
entry (that is, the first entry) with value 2. Therefore, \textit{Researcher} nodes range from $3$ to $5$.

\item \textbf{Find by element identifier}. $\textit{Get}\{\textit{Node}|\textit{Edge}\}\textit{Type}(\textit{id)}$ gets the type corresponding with an identifier. \textit{GetNodeType} starts by performing a binary search over the upper-limits of the Node Schema, until the correct range is found. Then the label can be returned. For instance,  $\textit{GetNodeType}(4)$ starts from a binary search over the nodes schema, until the lowest upper-limit is found (in this case is the second entry with value 5) and the highest lower limit (the first entry with value 2). Consequently, the node $4$ has type \textit{Researcher}. The behavior of $\textit{GetEdgeType}(\textit{id})$ is totally symmetric to $\textit{GetNodeType}(\textit{id})$ and it is implemented exactly in the same way.
\end{itemize}

\subsection{Operations over the Data}
Next four operations involve the data subsystem. They work over the attribute values of the nodes and edges of the graph.

\begin{itemize}

\item \textbf{Attribute retrieval}. $\textit{Get}\{\textit{Node}|\textit{Edge}\}\textit{Attribute}(\textit{id},\textit{att})$ is the basic operation that obtains the value that a node (or edge) with identifier \textit{id} takes for the attribute with label \textit{att}. The operation starts by obtaining the type of the given node, which is solved with the operation $\textit{GetNodeType}(\textit{id})$. Then, the list of valid attributes of the node is checked looking for label $att$. If label \textit{att} is not included in the list of valid attributes for that type, then the attribute is undefined for that node and no result is returned. For instance, $\textit{GetNodeAttribute}(3,``\textit{Title}")$ in the example searches the list of attributes of type \textit{Researcher}, which are \textit{Name}, \textit{University}, and \textit{Position}; thus, \textit{Title} is not a valid attribute and no result is returned. Otherwise, the attribute is checked. By accessing to the label bitmap it is possible to determine if the attribute is sparse or dense. If it is a sparse attribute, the procedure is quite simple: the list of plain values is checked at position $id-\textit{limit}+1$, where $\textit{limit}$ is the lowest identifier of the type $\textit{GetNodeType}(\textit{id})$. The value of this cell is returned. For instance, for $\textit{GetNodeAttribute}(3,``\textit{Name}")$ it obtains that \textit{Name} is the first label of node type \textit{Researcher}, thus, it checks the first position of its bitmap. It is a 0, thus, it is a sparse attribute. Then it will return the value of the position 2 in the list of values for \textit{Name}, that is, \textit{J. Boy}. For dense attributes, a range operation has to be performed in the $k^2$-tree. The range includes only one row (corresponding to \textit{id}) and the columns representing the dense attribute that is being checked. For instance, for the operation $\textit{GetEdgeAttribute}(6,``\textit{Expertise}")$ it obtains that \textit{Expertise} is the first label of edge type \textit{Reviewer}, thus, it checks the first position of its bitmap. It is a 1, thus, it is a dense attribute. Then, the row $6$ between the columns from $1$ to $3$ is checked. A \textit{one} appears in the second column, so the value corresponding to the second position in the \textit{att} list, \textit{Medium}, is finally returned as a value.

\item \textbf{Filter by attribute value}. $\textit{Select}\{\textit{Nodes}|\textit{Edges}\}(\textit{Type},\textit{att},\textit{val})$ returns all nodes (or edges) belonging to \textit{Type} that take value \textit{val} for attribute \textit{att}. It is a classical filtering by property and type. In the graph of the example, queries like \textit{researchers from Coru{\~n}a, papers with the topic Graph Compression} or \textit{PhD Students} are examples of this select operation. The operation starts by recovering the entry corresponding to the specified $Type$ in the same way $\textit{Scan}\{\textit{Nodes}|\textit{Edges}\}(\textit{Type})$ does. It obtains the lower and upper limits for the identifiers of that type, $(\textit{lower},\textit{upper})$, which will be necessary later for this query. Then, attribute $att$ is searched in the attribute list of that entry.  If it is a dense attribute, the value is searched in the list of labels of that attribute in order to compute the limits of the needed range search over the $k^2$-tree. For instance, when $\textit{SelectNodes}(``\textit{Researcher}",``\textit{Position}",``\textit{Chair}")$ is queried, a range query is performed between rows from $4$ to $5$ (since these are the lower and upper limits of the \textit{Researcher} type) and column 1 (corresponding to value \textit{Chair}). The rows taking a \textit{one} in this range will be returned as a result (in this case, nodes $3$ and $4$). Since attributes are located in the $k^2$-tree ordered by value, in addition to the equality, other patterns of comparison could be implemented efficiently. For sparse queries, the operation is similar (binary search over all the labels of this attribute list). When the valid values are reached, their positions determine the node identifiers which have to be returned.
\end{itemize}

\subsection{Operations over the Relationships}
The last kinds of queries involve conditions over the relationships of the graph. Two basic queries can be the basis of the exploration of the relationships in the graph:

\begin{itemize}
\item \textbf{Find neighbors by node type}. $\textit{Neighbors}(\textit{Type},\textit{id})$ returns all nodes of the specified type that are neighbors of the node with the identifier \textit{id}. The operation starts by retrieving the range of valid identifiers $(\textit{lower},\textit{upper})$ according to the given type. Then the multi-edge $k^2$-tree is explored in row $id$ and between columns $(\textit{lower},\textit{upper})$. Consider the query $\textit{Neighbors}(``\textit{Researcher}",4)$, asking for the neighbors of node 4 (\textit{researcher J. Boy}) which have \textit{Researcher} type.  First of all, the limits of \textit{Researcher} are computed, obtaining ($3,5$). Therefore, a range query between row $4$ and columns $(3,5)$ is performed. A multi-edge in the cell $(4,5)$ is recovered (containing edges $e_4$ and $e_5$); thus, the node $5$ (\textit{researcher S. G{\'o}mez}) is the result of that query.

\item \textbf{Find neighbors by edge type}. $\textit{Related}(\textit{Type},\textit{id})$ returns all nodes related to the identifier $id$ connected to them through an edge with the given \textit{Type}. In this operation, the filtering is in the edge identifier, which is recovered after performing the query over the $k^2$-tree. Hence, this filtering has to be processed after finishing the query. The query is executed as follows. First of all, the valid range of identifiers of the given edge type is computed. Therefore, the full row \textit{id} is queried in the multi-edge $k^2$-tree. After that, all the results are processed sequentially, removing from the result the columns that do not contain any edge in the range of edge identifiers for the given edge type. The result will be the identifiers of the remaining columns. For instance, the query $\textit{Related}(``\textit{Author}",3)$ starts  by computing the valid identifiers for $Author$, which are 1--3. Then, the row $3$ is queried in the multi-edge $k^2$-tree, obtaining two cells with results: $(3,1)$ and $(3,2)$. The edge identifier of the cell $(3,1)$, that is $e_1$, is included in the range of valid identifiers, so $n_1$ is a result of that query (the paper \textit{Compressing graphs}). However, the edge identifier in cell $(3,2)$ is not valid ($e_6$ has type \textit{Reviewer}) so node $n_2$ is not returned as a result.

\end{itemize}

The set of operations we implement in Att$K^2$-tree aims to provide a basic but efficient querying to the attributed graphs. More complex queries can be implemented on the top of these basic operations as intersections, unions or chains of them. For instance, the query \textit{Papers reviewed by P. Garc{\'i}a and written by S. G{\'o}mez} could be implemented as an intersection of three different operations:
\begin{itemize}
\item {\em ScanNodes(``Paper")}
\item  {\em Related(SelectNodes(``Reviewer", ``Researcher", ``Name", ``P.~Garc{\'i}a"))}, and
\item  {\em Related(SelectNodes(``Author", ``Researcher", ``Name", ``S.~G{\'o}mez"))}. 
\end{itemize}
The experimental evaluation in Section \ref{sec:GeneralEvaluation} gives some experimental results of the spatial requirements and the temporal efficiency obtained by Att$K^2$-tree. Furthermore, as a proof of concept, it is evaluated with other proposals in the state of the art.

\section{Dynamic AttK$^2$-tree}\label{sec:dynamic}
The approach described in the previous section is static, thus, it requires knowing in advance the whole graph database we want to store. However, this may be a limitation for some possible application scenarios. In this section, we describe the dynamic Att$K^2$-tree, denoted dynAtt$K^2$-tree, which allows changes in schema, data, and relationships.

\subsection{Data structure}
The dynAtt$K^2$-tree can also store a directed, attributed, and labeled multi-graph. Its dynamic nature relies on several dynamic data structures \cite[Chapter~12]{Nav16}, especially dynamic $k^2$-trees \cite{k2dyn}, dynamic wavelet trees \cite{MNtalg08}, dynamic bit arrays and dynamic vectors. Analogously to Att$K^2$-tree, dynAtt$K^2$-tree  represents the graph with three components: the schema of the data, the data included in the nodes and the edges and, finally, the relations between the elements of the graph topology. Next, we present these three components.

\subsubsection{Schema}

Whereas the static Att$K^2$-tree reorders nodes and edges by type, such that by knowing their identifier, their type can be efficiently computed (by means of a binary search), this cannot be done in the dynamic variant. Node and edge identifiers are given in order as they are inserted into the database. Thus, we use two dynamic sequences to represent the node/edge types, sorted by their identifier. More concretely:

\begin{itemize}

\item \textbf{Nodes schema}: the nodes schema keeps the valid node labels, and the label each node of the graph has. To keep this part compact and dynamic, node labels are stored using dynamic vectors where the labels are sorted lexicographically. In addition, the type of each node is stored using a dynamic sequence, more concretely using dynamic wavelet trees, which allow us to efficiently locate all nodes of a given type and also obtain the type of a given node, using little space\footnote{A wavelet tree is a data structure that maintains a sequence $S$ of $n$ symbols
supporting the following operations: $access(S, i)$, which returns the symbol at position $i$ in $S$; $rank_c(S, i)$, which counts the times symbol $c$ appears up to position $i$ in $S$; and $select_c(S, j)$, which returns
the position in $S$ of the $j$-th appearance of symbol $c$. They can be efficiently implemented using compressed space and perform well in practice. Wavelet trees and their applications have been extensively described by \citeasnoun{Navjda13}.}. For each node label, we include a vector of attributes and a dynamic bit array, indicating, for each attribute, if it is a sparse or dense.

\item \textbf{Edges schema}: the edges schema is implemented in the same way as the nodes schema, storing the edge types sorted lexicographically in a dynamic vector. Again, dynamic wavelet tree is used for storing the type of each edge in compact space while still allowing efficient searches. Again, for each edge label, we include a vector of attributes and a dynamic bit array, indicating, for each attribute, if it is a sparse or dense.

\end{itemize}

As with the static version, the schema layer is the starting point of the internal representation of the graph in the dynAtt$K^2$-tree, providing indexed access to the other two layers. It is used to retrieve the different labels of node and edges, and to recover the label for a given identifier. It also stores references to the valid properties for a given label. In contrast to the static version, the proposed dynamic variant allows changes on the node or edge schema. For instance, it is possible to include new node/edge types or new attributes for a node/edge type.

\subsubsection{Data}

We also differentiate among sparse and dense attributes:

\begin{itemize}
	
	\item \textbf{Sparse attributes}: are represented analogously to the sparse attributes in the static version, but using dynamic lists.
	
	\item \textbf{Dense attributes}: are represented slightly different compared to the static version. Att$K^2$-tree uses only two $k^2$-trees, one for the dense nodes attributes and another for the the dense edges attributes. We allow changes in the data schema, and more particularly, new value attributes can appear for dense attributes. Since the dynamic $k^2$-tree does not support efficiently adding columns or rows in the middle of the matrix, but at the end, it becomes more convenient to use one different dynamic $k^2$-tree for each dense attribute, such that new attribute values are always appended at the end of its attribute matrix. 
	
\end{itemize}

\subsubsection{Relations}
The third component of dynAtt$K^2$-tree follows the same approach as the static Att$K^2$-tree. It uses a dynamic $k^2$-tree for storing the relations among nodes and a dynamic bit array to store bitmap $\textit{Multi}$. The main difference is found in the representation of the list of multiedges for a given pair of nodes. In the dynamic variant, we do not use arrays $\textit{Last}$ and $\textit{More}$, but a dynamic vector containing, for each leaf of the tree, a dynamic vector of edge identifiers.

\subsection{Operations and Navigation}
Navigation is done analogously to that of the static Att$K^2$-tree, described in Section \ref{subsec:nav}. 
The main substantial difference appears for the operations over the schema $\textit{Get}\{\textit{Node}|\textit{Edge}\}\textit{Type}(\textit{id})$, as labels do not depend on their identifier, but must be queried over a wavelet tree. 

Moreover, as this dynamic version allows changes in the schema, the data, and the relationships (insertions, deletions and updates), new operations appear to support these functionalities. Basically, these operations rely on the dynamism of the underlying structures (insertions, deletions and updates over the dynamic $k^2$-trees, dynamic wavelet trees, dynamic bit arrays and dynamic vectors).

\section{Experimental evaluation}\label{sec:GeneralEvaluation}

In this section, we analyze the spatial and temporal performance of our structure, which was designed to support basic operations over an attributed graph in a very compact way.
We compare our structure with DEX, Neo4j, and HANA Graph, three of the most relevant graph database management systems in the state of the art. However, it is important to note that the results are provided in order to prove that we propose a compact structure with some basic search capabilities that is competitive in terms of space and time, but we are not proposing an alternative to these systems, since the purposes of our structure are different. We designed a compact attributed graph representation with some queryable capabilities and we implemented some basic operations, but our structure is not a full graph database platform. Neither Att$K^2$-tree nor dynAtt$K^2$-tree support all the algorithms and operations that are characteristic in those kinds of engines. Thus, this comparison has to be understood just as a proof of concept of the structure we propose.

\subsection{Experimental Framework}
We ran experiments on a dedicated Intel\textsuperscript{\textregistered} Core\textsuperscript{TM} i7-8700K CPU @ 3.70GHz (12 cores) with 12MB of cache, and 64GB of RAM. It ran Ubuntu 16.04.1 LTS with kernel 4.4.0-31 (64 bits).

\subsubsection{Tools} \label{subsec:toolsAtt}

We include here some specific details on how each system was configured for launching the queries. 

\begin{itemize}
\item Att$K^2$-tree: was implemented in C, compiled with gcc (version 5.4.0).
\item dynAtt$K^2$-tree: was implemented in C++, compiled with gcc (version 5.4.0). It uses 
{\em DYNAMIC}\footnote{\url{https://github.com/xxsds/DYNAMIC}}, a succinct and compressed dynamic data structures library implemented by Nicola Prezza  \cite{prezza}.
\item DEX: corresponds to the very compact graph database described in Section \ref{sec:DEX}. We implemented the operations that Att$K^2$-tree supports through a Java program that invokes the corresponding native functions of the DEX library.
\item Neo4j: is the commercial graph database system described in Section \ref{sec:neo4j}. In order to execute the same operations implemented over Att$K^2$-tree, the queries were implemented in Cypher language. Those Cypher queries are called from a program implemented in Java, which uses the Neo4j Java driver. 
\item HANA Graph: is the commercial graph database system described in Section \ref{sec:hana}. More concretely, we have used the official SAP HANA, express edition, which is available for free. We implemented the queries using a program implemented in Java and connecting to the database using a JDBC driver, more concretely using the ngdbc library (sap.jdbc).
\end{itemize}

\subsubsection{Queries} \label{subsec:queriesAtt}
We measured the performance of our structure through the execution of $8$ different kinds of queries, whose implementation in our structure was described in Section \ref{subsec:nav}. They include operations over the schema, the relations and the attributes of the graphs.\\

We designed a synthetic query set of 1,000 queries per each kind of operation:

\begin{itemize}

\item \textbf{Query set 1}: \textit{GetNodeType} obtains the type of a given node.
\item \textbf{Query set 2}: \textit{GetEdgeType} obtains the type of a given edge.
\item \textbf{Query set 3}: \textit{GetNodeAttribute} obtains the value that a given node takes in a specific attribute.
\item \textbf{Query set 4}: \textit{GetEdgeAttribute} obtains the value that a given edge takes in a specific attribute.
\item \textbf{Query set 5}: \textit{SelectNode} obtains the set of nodes that takes a given value for an attribute.
\item \textbf{Query set 6}: \textit{SelectEdge} obtains the set of edges that takes a given value for an attribute.
\item \textbf{Query set 7}: \textit{Neighbors} returns the nodes of a given type related to a node.
\item \textbf{Query set 8}: \textit{Related} returns the nodes related to a given one through a specific edge type.

\end{itemize}

Note that the previously described operations \textit{ScanNodes} and \textit{ScanEdges} are not included in this evaluation as they are relevant only for our proposed structure. We have not included \textit{GetNodeTypes} and \textit{GetEdgeTypes}either, as they can be executed very fast and they lack of interest in this comparison.\\

These query sets are analyzed in three categories. On one hand, the operations over the schema (queries 1 and 2), which are the most simple queries. The times obtained by each alternative for these operations give a brief idea of the minimal time of communication with the database. Queries from 3 to 6 represent operations over the data (properties and types) of nodes and edges. Finally, query sets 7 and 8 establish conditions over the relationships in the graph.

\subsubsection{Datasets}

\begin{figure}
  \centering
      \includegraphics[width=0.6\textwidth]{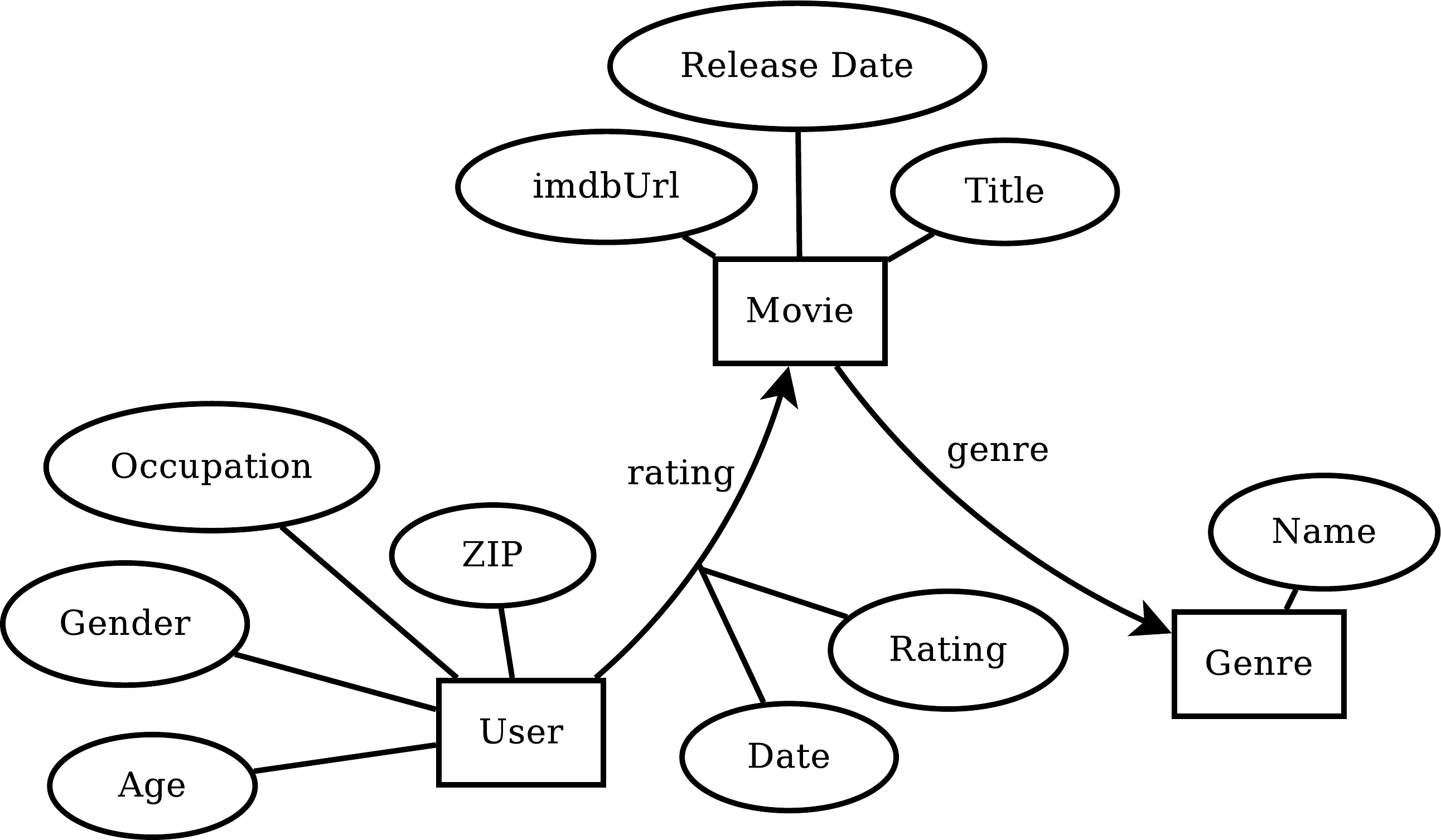}\\
       \vspace{1cm}
      \includegraphics[width=0.57\textwidth]{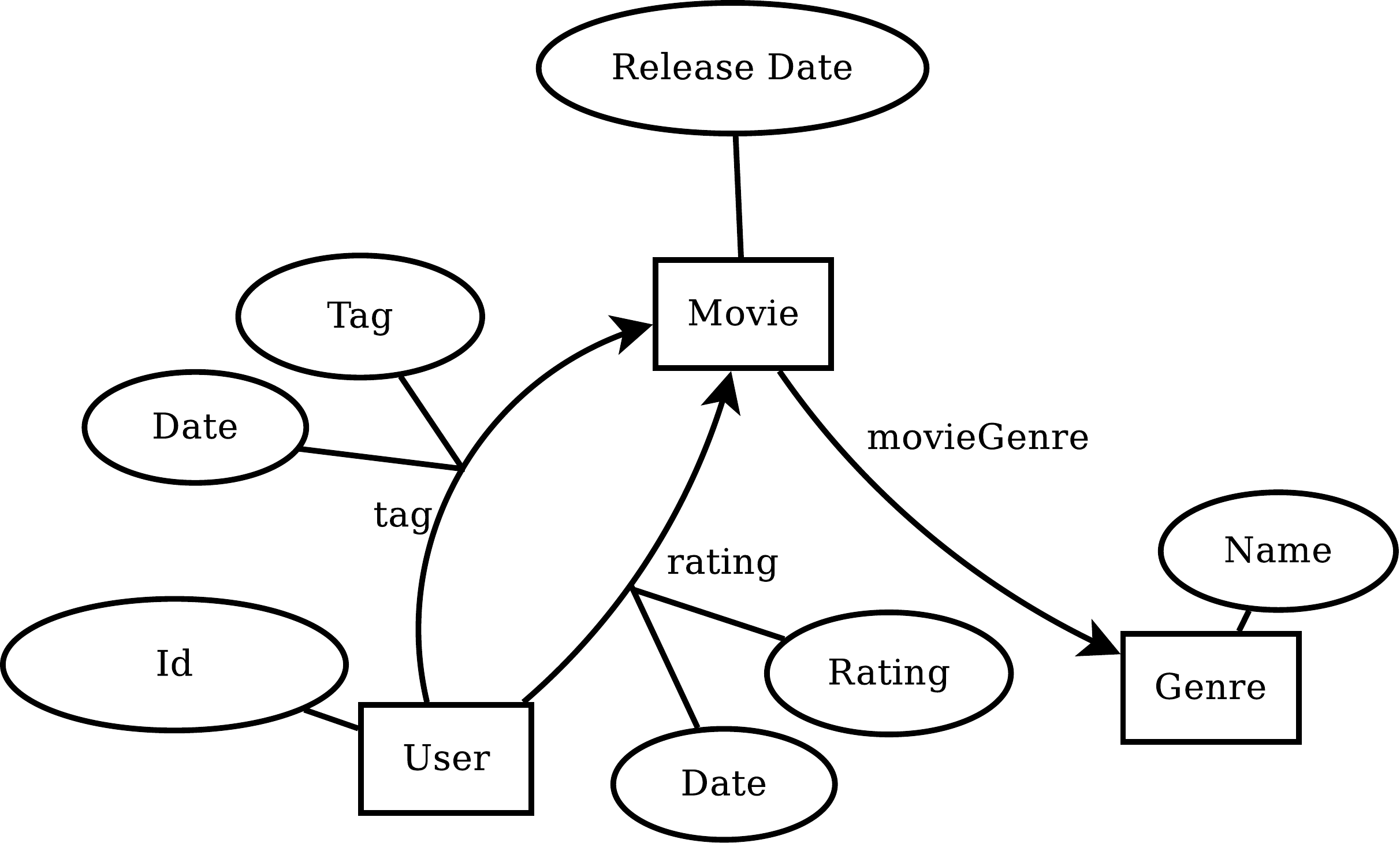}\\
       \vspace{1cm}
      \includegraphics[width=0.7\textwidth]{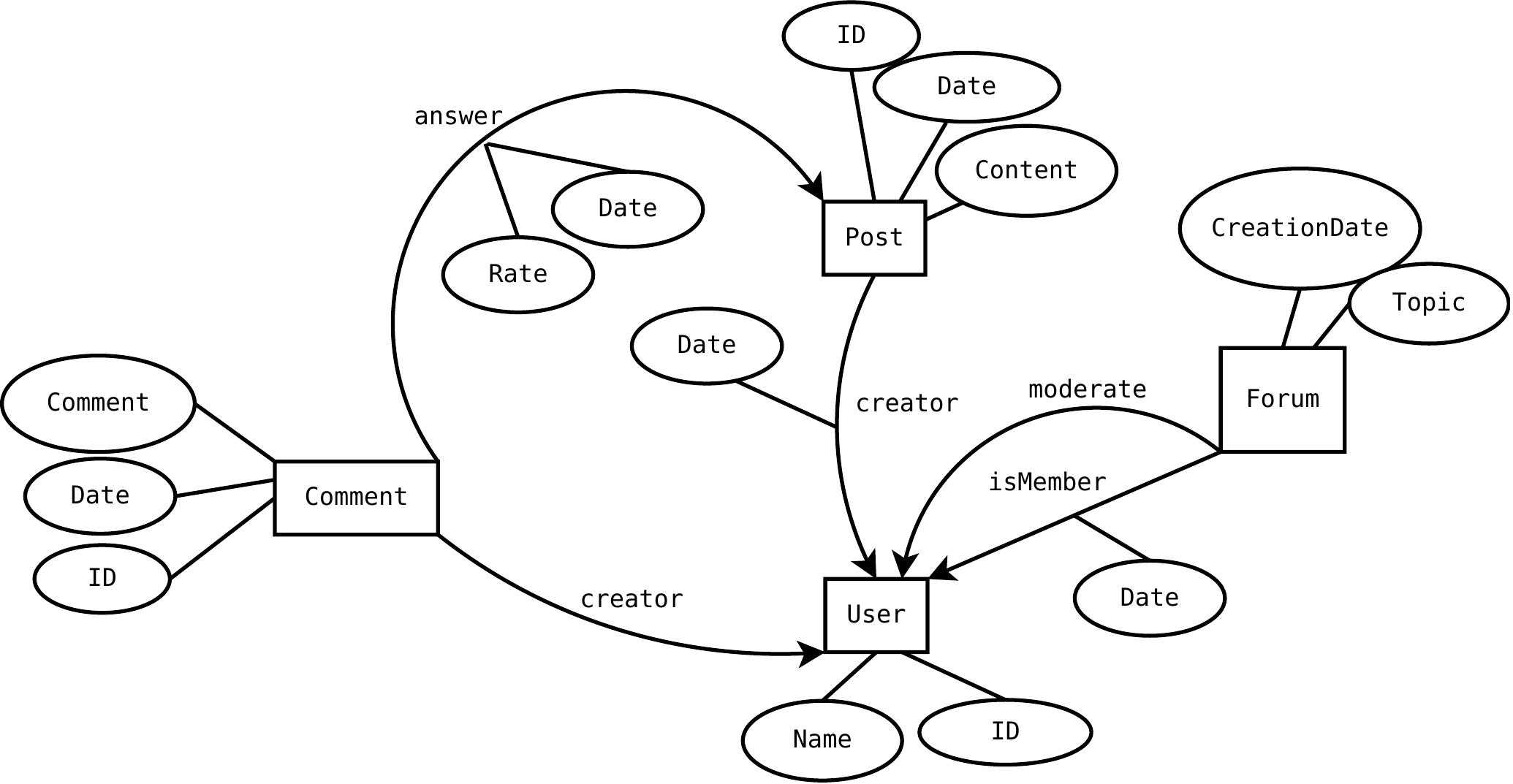}
  \caption{Attributed graph representing \texttt{ML100k} (top), \texttt{ML10m} (center), and \texttt{SNBsmall} and \texttt{SNBlarge} (bottom) datasets.}
  \label{fig:datasetmodels}
\end{figure}

\paragraph*{Movielens 100k (\texttt{ML100k})}

The first use case we analyze is a dataset extracted from a movie recommendation website, Movielens\footnote{http://movielens.umn.edu/}, which contains ratings of movies from different users of the web, including statistical information of the users and tags of the movies. We use a subset of 100,000 ratings for 1,682 movies from 943 users. It also contains 2,874 movie-genre associations for 19 possible genres \cite{MOVIELENS}.

Figure \ref{fig:datasetmodels} (top) shows the attributed graph representing the small movielens dataset, which we will represent using DEX, Neo4j, HANA Graph and our own structures. The graph model for this dataset has three kinds of entities: users, movies and genres. Movies and users contain attributes presenting different value distributions. Regarding to the representation in our structure, we  use a $k^2$-tree to represent the dense attributes $Age$, $Gender$ or $Occupation$, while the remaining sparse attributes are directly stored through an indexed-plain list.

\paragraph*{Movielens 10m (\texttt{ML10m})}
We analyze the results of a different dataset that also represents movie recommendations from Movielens. The model, which is shown in Figure \ref{fig:datasetmodels} (center), contains a more reduced set of properties. However, the number of entities is larger than the previous use case. This dataset contains  10,000,053 ratings for 10,681 movies from 71,567 users \cite{MOVIELENS}, where more than 20 million of properties need to be stored. It also contains 21,564 movie-genre associations for 20 possible genres, and 95,581 tags given by users to movies.

\paragraph*{LDBC-SNB (\texttt{SNBsmall} and \texttt{SNBlarge})}
We also create two synthetic datasets using the LDBC Social Network Benchmark \cite{LDBCSNB} (LDBC-SNB). The model, which is shown in Figure \ref{fig:datasetmodels} (bottom), simulates a realistic social network containing three type nodes (Users, Posts, Forums, and Comments) and several edge types (answer, creator, isMember, moderator). We tune the configuration files to vary cardinalities and distributions, thus generating two graph datasets of different size, \texttt{SNBsmall} and \texttt{SNBlarge}. The small variant contains  336,428 nodes (184,091 comments, 16,256 forums, 4,515 people, 131,566 posts) and 848,873 edges from different types. The large variant contains  10,547,201 nodes (7,329,735 comments, 249,935 forums, 27,007 people, and 294,0524 posts) and 26,700,489 edges from different types.
Regarding the representation in our structure, we use a $k^2$-tree to represent the dense attributes \textit{Topic} and \textit{Date} at the nodes, and the attribute \textit{Rate} for edges of type {\em Answer}, while the remaining attributes are considered sparse and are directly stored through an indexed-plain list.

\subsection{Results}

\begin{figure}
  \centering
      \includegraphics[width=0.8\textwidth]{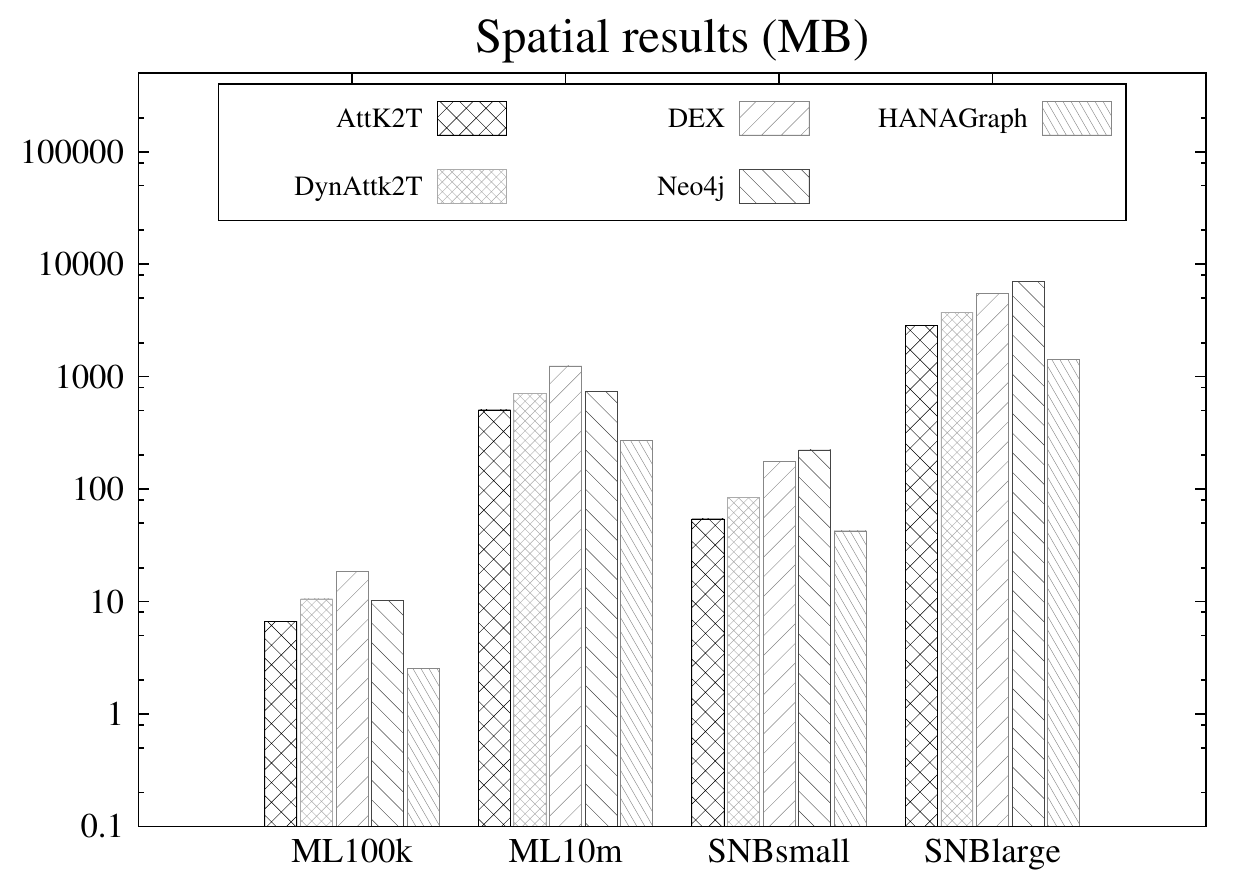}
  \caption{Spatial results obtained by each system over the four datasets.}
  \label{fig:space}
\end{figure}

We first show the spatial cost for representing all datasets for the five different approaches. Figure \ref{fig:space} shows the cost in megabytes. Note that in the case of Att$K^2$-tree and dynAtt$K^2$-tree, we show the memory usage in main memory, while in the case of Dex and Neo4j the results are measured as their cost in secondary memory with the graph engine system offline. For HANA Graph, we use the estimated maximum memory consumption data reported in the table run information.  
Our proposal achieves better spatial results than DEX and Neo4j. 
Compared to HANA Graph, we obtain slightly larger sizes. However, we will see in the following temporal comparisons that this solution is much worse in terms of time performance. \\

\begin{figure}
  \centering
      \includegraphics[width=0.9\textwidth]{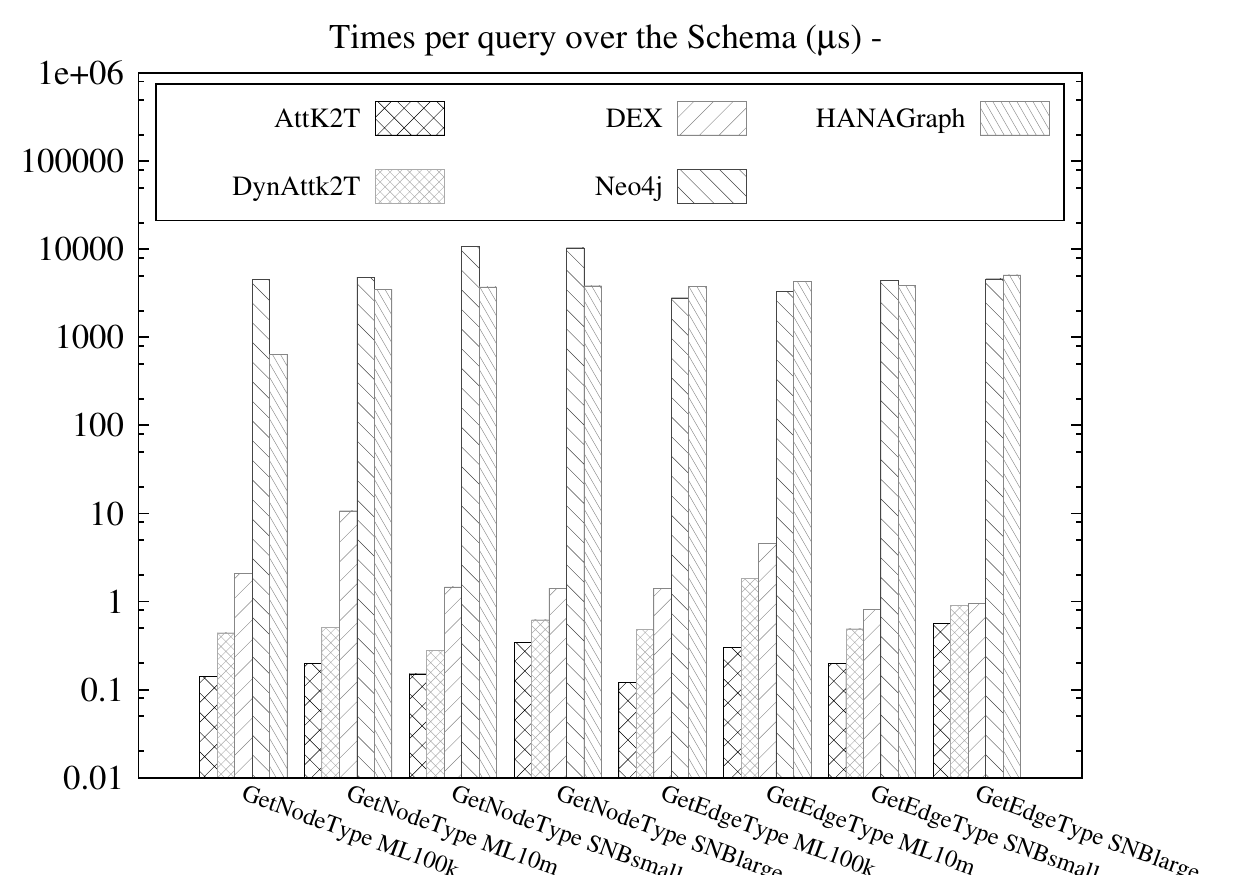}
  \caption{Time results obtained for operations over the schema: {\em GetNodeType, GetEdgeType}.}
  \label{fig:timesschema}
\end{figure}

Figure \ref{fig:timesschema} shows some temporal results of simple and fast operations over the schema of the graph. More concretely, we ask for the type of a node or an edge (queries \textit{GetNodeType} and \textit{GetEdgeType}). They are very lightweight operations in our system, particularly for Att$K^2$-tree, since each type has a range of identifiers. Thus, given an identifier, we only need to search over the list of node or edge types, which usually contains very few elements. In the case of dynAtt$K^2$-tree, this operation is solved using only an access operation over a dynamic wavelet tree, which is efficient in practice. Only DEX obtains similar results to the dynamic version when solving \textit{GetEdgeType} over the largest dataset (\texttt{SNBlarge}). It is important to mention that in the case of DEX, Neo4j, or HANA Graph, as we described in Section \ref{subsec:toolsAtt}, every query performed in our experimental evaluation involves the parsing of the query, the connection with the database and other operations.\\ 

\begin{figure}
  \centering
      \includegraphics[width=0.9\textwidth]{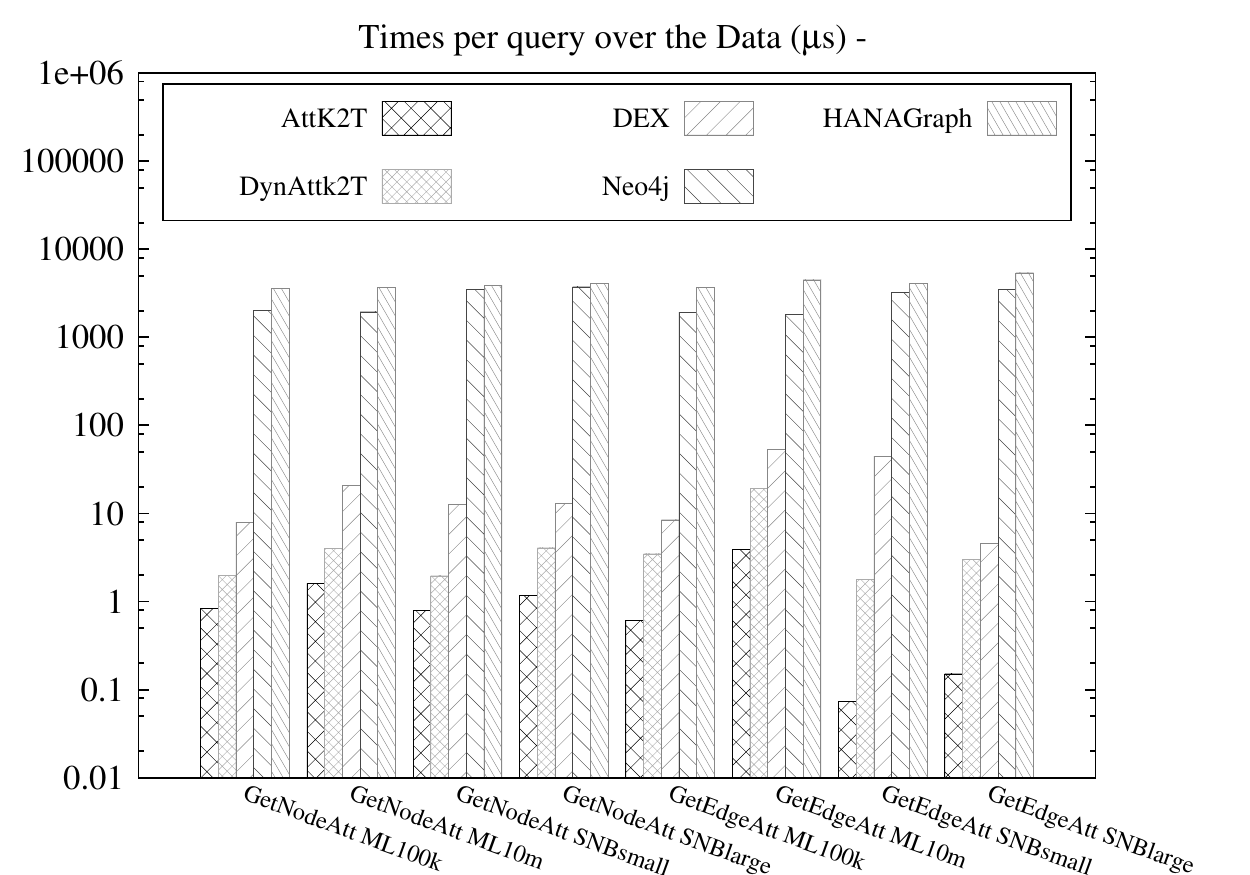}\\
      \vspace{1cm}
      \includegraphics[width=0.9\textwidth]{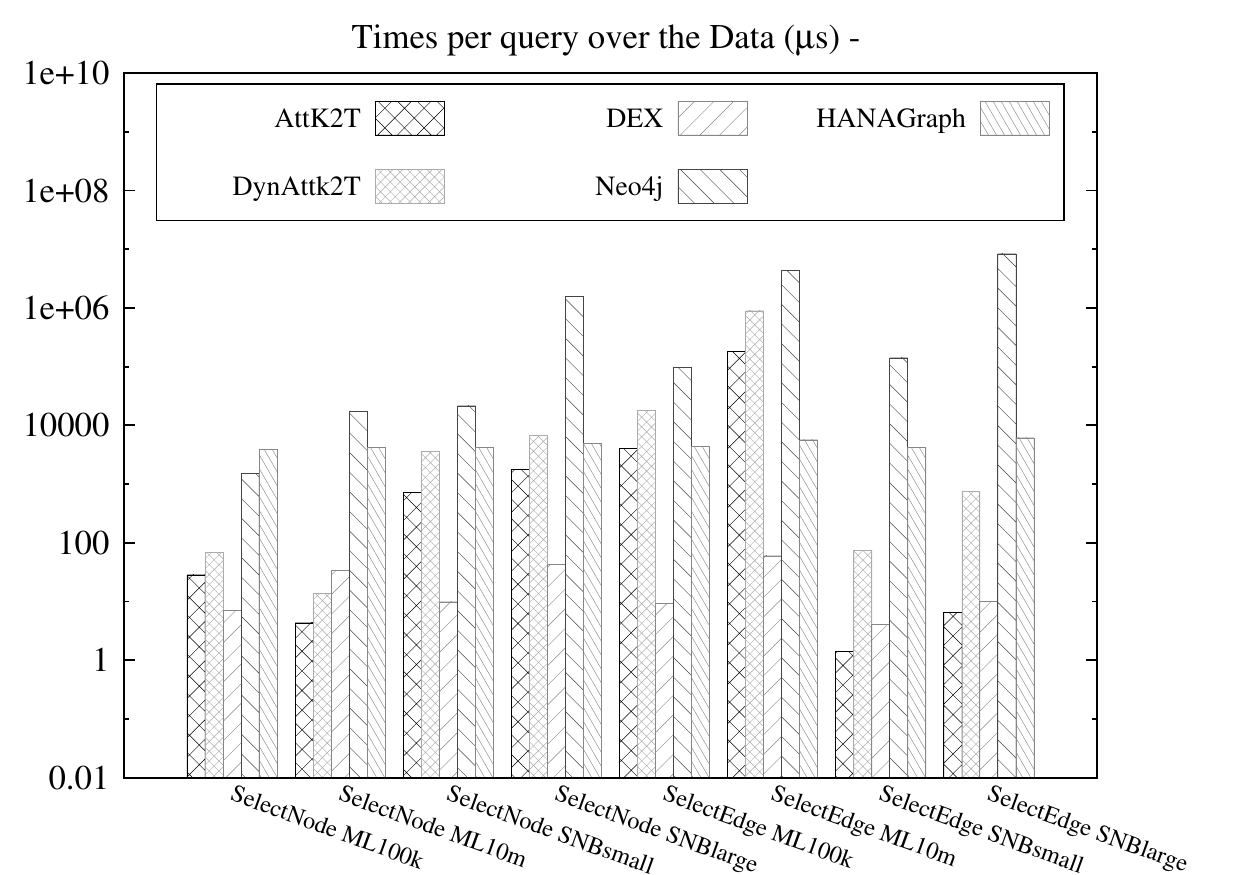}
  \caption{Time results obtained for operations over the data: {\em GetNodeAttribute, GetEdgeAttribute}(top) and {\em SelectNode, SelectEdge} (bottom).}
  \label{fig:movielensdata}
\end{figure}

Figure \ref{fig:movielensdata} shows the results obtained for the operations over the graph data. In both of our variants, Att$K^2$-tree and dynAtt$K^2$-tree, we can see how obtaining the property value for a given node/edge (operations \textit{GetNodeAttribute} and \textit{GetEdgeAttribute}) is faster than obtaining the list of nodes/edges that take a given value (operations \textit{SelectNode} and \textit{SelectEdgee}). Compared with the other systems, our proposals are faster for \textit{GetNodeAttribute} and \textit{GetEdgeAttribute}. However, they are outperformed by DEX when solving \textit{SelectNode} and \textit{SelectEdge} over most of the datasets. Neo4j and HANA Graph are slower for all the operations. Att$K^2$-tree and dynAtt$K^2$-tree obtain better results than DEX for dataset \texttt{ML10m}. This is due to the fact that all node attributes are sparse, and they are faster to retrieve than dense attributes for our proposal. This happens also with edge attributes for the synthetic datasets (\texttt{SNBsmall} and \texttt{SNBlarge}), as most of them are stored as sparse attributes; thus, Att$K^2$-tree  obtains better results than DEX when performing \textit{SelectEdge}.\\ 

\begin{figure}
  \centering
      \includegraphics[width=0.9\textwidth]{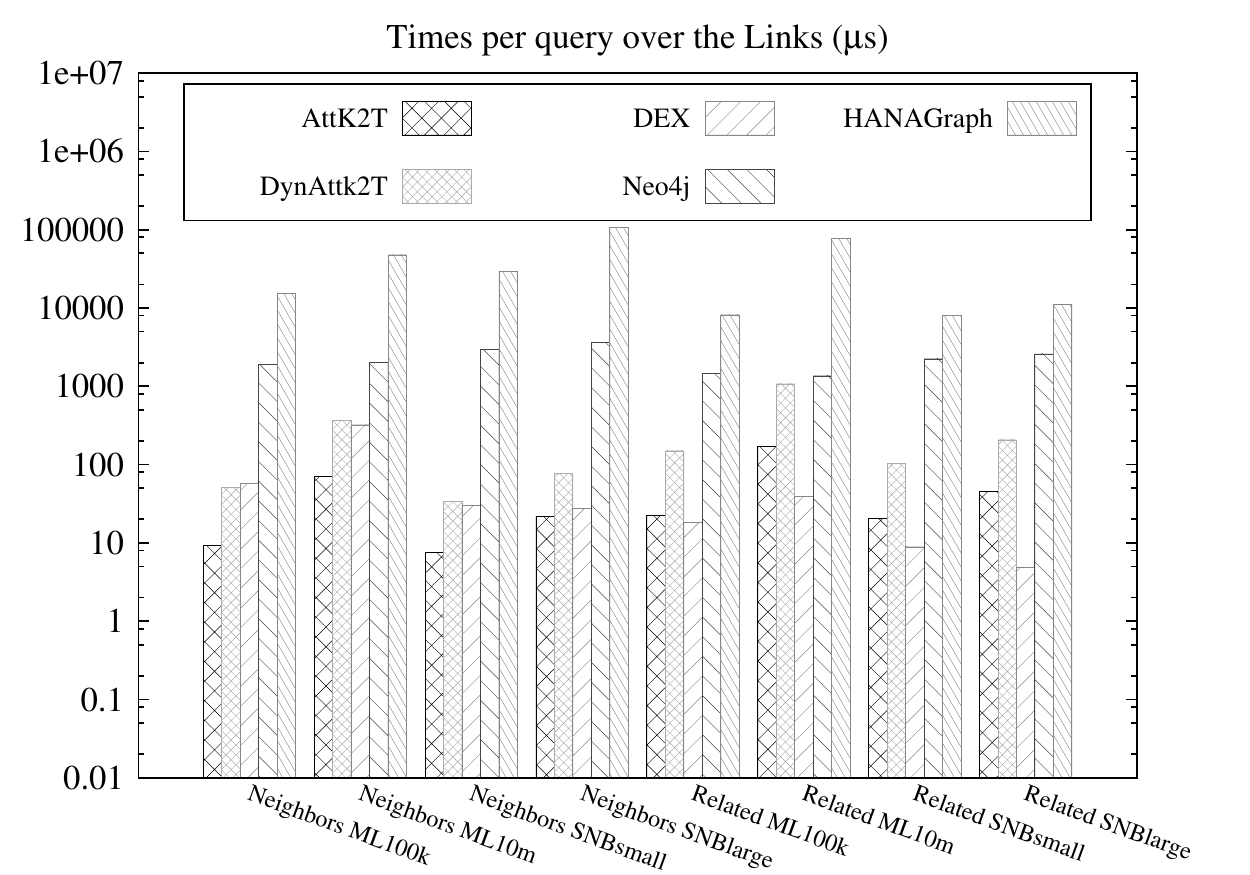}
  \caption{Time results obtained for operations over the links: {\em Neighbors, Related}.}
  \label{fig:movielenslinks}
\end{figure}

Finally, Figure \ref{fig:movielenslinks} shows operations over the relations among nodes. Links in Att$K^2$-tree and dynAtt$K^2$-tree are stored using a $k^2$-tree (and an extra structure of bitmaps), so these operations are very fast in our structures, since they do not involve operations over the attributes. Operation \textit{Related} is slower than \textit{Neighbors} for Att$K^2$-tree and dynAtt$K^2$-tree, because \textit{Related} implies an additional filtering of the final list of candidate edges, which is not necessary in the case of the \textit{Neighbors} operation. DEX system improves the results obtained by our proposal, while Neo4j and HANA Graph are slower in both operations for all datasets.

\subsection{Summary of results}
Att$K^2$-tree is a very compressed representation of attributed graphs that supports efficient access to the properties and the relationships of the elements of the graph. The structure we propose is not a full graph database engine, as we only support a set of basic queries. However, the spatial and temporal results obtained in the experimental evaluation show that it is a very competitive approach to represent static graph data in a very compact way and to perform basic graph operations over the compressed structure. The dynamic variant achieves worse space/time results compared to the static alternative, as expected. However, compared to the rest of the graph database systems, which also allow flexible schemas and modifications of the data, it obtains a very good compromise between space requirements and query performance.

\section{Conclusions and future work}
We presented a new compact representation of attributed graphs that supports efficient access to the nodes, edges and their properties. More concretely, our proposal is designed for representing and navigating labeled, directed, attributed multigraphs in little space and efficient time. It works in main memory and it relies on the $k^2$-tree structure for representing most of the data. Relations among nodes are represented using an extension of the $k^2$-tree that supports multiple edges among the same pairs of nodes. Regarding to the properties of the elements, we differentiate among dense attributes (presenting very few different values), which are stored using $k^2$-trees, and sparse attributes, stored as plain lists.

We presented two different variants of the representation: a static version, denoted Att$K^2$-tree, and a dynamic version, denoted dynAtt$K^2$-tree. We experimentally evaluated the spatial and temporal performance of both of our variants for datasets of different nature and size. We also stored the same data in DEX, Neo4j, and HANA Graph in order to provide some spatial and temporal references of other systems. Results showed that our proposals obtain competitive space and time results, compared with these existing graph database management systems. However, it is important to note that these systems are full attributed graph engines with many features and possible combinations, in addition to complete query APIs, so this comparison has to be understood only as a proof of concept of our structure.

The $k^2$-tree is the basis of the proposed approach, as it is used for representing the binary relations among nodes and also among nodes/edges with attribute values. The compact spaces obtained by our solution are due to the good properties of $k^2$-trees in terms of space. In any case, our solution can be regarded as a modular system, where each part can be replaced with other compact data structures that improve the space/time trade-off. For instance, instead of $k^2$-trees, one could explore other promising representations that have appeared recently for graph compression \cite{GLOUDS,Maneth2016}, as soon as they become mature enough to compete with $k^2$-tree not only in terms of space, but also in terms of scalability and extended functionality. 

We implemented a basic set of operations to query the properties and the connections of the elements of the graph. A future line of research will include the design and implementation of algorithms to solve more complex operations. 
In addition, we will explore the influence of node and edge reordering in our proposal. As studied by \citeasnoun{BLN14}, node ordering is a key aspect for obtaining high compression in the $k^2$-tree structure. In the case of Att$K^2$-tree, nodes and edges are sorted according to their labels, in a decision made to save space and improve performance for some queries. This is not further used in the dynamic variant, dynAtt$K^2$-tree, as node and edge identifiers are given sequentially as they are inserted. Thus, using the same data structures as the dynamic variant for storing node/edge types, we can reorder node/edge identifiers in the static version, trying to minimize space consumption.

\section*{Acknowledgements}
This research has received funding from the European Union's Horizon
2020 research and innovation programme under the Marie Sk{\l}odowska-Curie
[grant agreement No 690941]; from the Ministerio de  Econom\'{\i}a y Competitividad (PGE and ERDF) [grant numbers TIN2015-69951-R; TIN2016-77158-C4-3-R] and from Xunta de Galicia (co-founded with ERDF) [grant numbers ED431C 2017/58;
ED431G/01]. We also thank Nieves R. Brisaboa for her contributions during the initial discussions of this work.

\bibliographystyle{agsm}
\renewcommand\harvardand{and}
\bibliography{paper}

\section*{Author Biographies}
\leavevmode

\vbox{%
	\begin{wrapfigure}{l}{80pt}
		{\vspace{-10pt}\includegraphics[width=70pt]{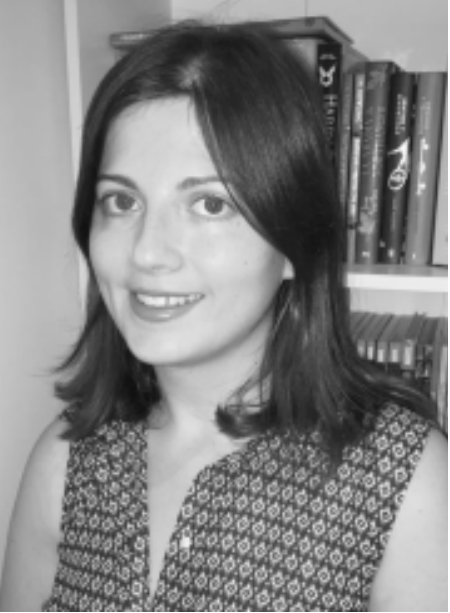}}%
	\end{wrapfigure}
	\noindent\small 
	{\bf Sandra {\'A}lvarez-Garc{\'i}a} is a software analyst and engineer at Indra and a collaborating researcher of the Database Laboratory at the University of A Coru{\~n}a. She received her degree in Computer Science Engineering in 2009 and her Ph.D. degree in Computer Science in 2014, both from the University of A Coru{\~n}a. 
	Her research was mainly focused on obtaining compressed and efficient representation of graphs, and more particularly, for managing RDF and linked data. Part of her research was carried out at Yahoo! Labs Barcelona and Yahoo! Labs Santiago de Chile.}

\vspace{40pt}
\vbox{%
	\begin{wrapfigure}{l}{80pt}
	{\vspace{-10pt}\includegraphics[width=70pt]{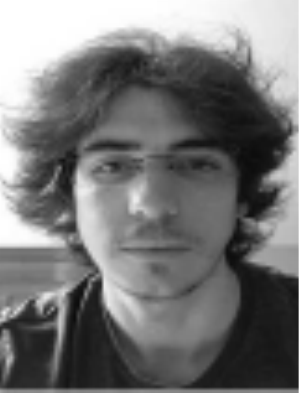}}%
	\end{wrapfigure}
	\noindent\small {\bf Borja Freire} obtained his degree in Computer Science at the University of A Coru{\~n}a in 2016. During his studies, he was awarded with a collaboration grant in the Department of Computer Science.
	In 2018, he obtained his Master degree in Bioinformatics at the same university. At the same time that he finished his master studies, he was hired by Enxenio S.L. to work in R\&D projects related to Bioinformatics. In addition, he was accepted into the
	Doctorate Program in Computer Science at University of A Coru{\~n}a, and he has been a Ph.D. student since then.
}

\vspace{40pt}
\vbox{%
	\begin{wrapfigure}{l}{80pt}
	{\vspace{-10pt}\centering \includegraphics[width=70pt]{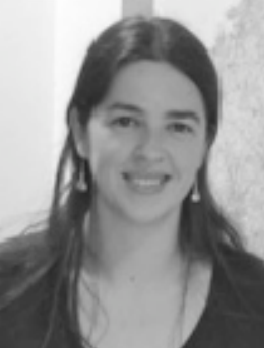}}%
	\end{wrapfigure}
	\noindent\small 
	{\bf Susana Ladra}  is Associate Professor at the University
	of A Coru\~na, where she obtained her degree in Computer Science Engineering in 2007 and her Ph.D. degree in Computer Science in 2011 at the same university. She also received her Bachelor in Mathematics from the
	National Distance Education University (UNED)
	in 2014. Her fields of interests include the design and analysis of algorithms and data structures,
	data compression and data mining in the fields of information retrieval and bioinformatics.
	She has published more than 40 papers in various international journals and conferences and is principal investigator of several national and international research projects.}

\vspace{30pt}
\vbox{%
	\begin{wrapfigure}{l}{80pt}
		{\vspace{-10pt}\centering \includegraphics[width=70pt]{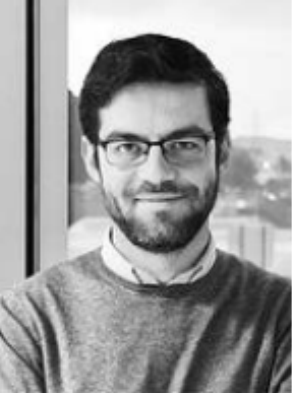}}%
	\end{wrapfigure}
	\noindent\small 
	{\bf {\'O}scar Pedreira} has M.Sc. and Ph.D. degrees in Computer Science from University of A Coru{\~n}a. He is an Associate Professor since 2008 at the same institution. He is a researcher of the Database Laboratory. His research interests include algorithms for similarity search, data structures and algorithms for graph databases, geographic information systems, and software engineering. He has co-authored many articles published in journals and conferences relevant for the research areas mentioned. He has continuously participated in research projects and technology and knowledge transfer projects with different companies.}

\vspace{40pt}

\correspond{Susana Ladra, Universidade da Coru{\~n}a, CITIC, Database Laboratory, Campus de Elvi{\~n}a, 15071, A Coru{\~n}a, Spain. Email: sladra@udc.es}
\label{lastpage}

\end{document}